\documentclass[aip,jap,reprint,groupaddress,noshowpacs]{revtex4-1}
\pdfoutput=1

\usepackage[colorlinks=false,bookmarks=true]{hyper ref}
 
\usepackage{mathrsfs}
\usepackage{epsfig}
\usepackage{amsmath}
\usepackage{subfigure}
\usepackage{amsfonts}
\usepackage{lineno}

\begin{document}

%
%

\title{Rock physics and geophysics for unconventional resource, multi-component seismic, quantitative interpretation}
%

%
%

\author{Michael E. Glinsky}
\affiliation{ION Geophysical, Houston, TX, USA}
\author{Andrea Cortis}
\affiliation{ION Geophysical, Houston, TX, USA}
\author{Jinsong Chen}
\affiliation{Lawrence Berkeley National Laboratory, Berkeley, CA, USA}
\author{Doug Sassen}
\affiliation{ION Geophysical, Houston, TX, USA}
\author{Howard Rael}
\affiliation{ION Geophysical, Houston, TX, USA}

%
%

\begin{abstract}
An extension of a previously developed rock physics model is made that quantifies the relationship between the ductile fraction of a brittle/ductile binary mixture and the isotropic seismic reflection response.  By making a weak scattering (Born) approximation and plane wave (eikonal) approximation, with a subsequent ordering according to the angle of incidence, singular value decomposition analysis are done to understand the stack weightings, number of stacks, and the type of stacks that will optimally estimate the two fundamental rock physics parameters.  Through this angle ordering, it is found that effective wavelets can be used for the stacks up to second order.  Finally, it is concluded that the full PP stack and the ``full'' PS stack are the two optimal stacks needed to estimate the two rock physics parameters.  They dominate over both the second order AVO ``gradient'' stack and the higher order (4th order) PP stack (even at large angles of incidence).  Using this result and model based Bayesian inversion, the detectability of the ductile fraction (shown by others to be the important quantity for the geomechanical response of unconventional reservoir fracking) is demonstrated on a model characteristic of the Marcellus shale play.
\end{abstract}

%
%

\maketitle

%
%

\section{Introduction}

The developing commercial significance of unconventional shale reservoirs is leading to the need to be able to remotely determine the ability to effectively fracture the reservoir.  This paper will establish the theory and practicality of optimally estimating the ductile fraction from an isotropic analysis of surface conventional and converted wave seismic data.  This property of a binary ductile/brittle mixture has been shown to be the key property in determining the geomechanical fracturing response of an unconventional reservoir \citep{zobak.etal.12,kohli.etal.13}.  This is, most likely, because of the balance between the ``bumpy road'' friction of the fracture, due to the structurally competent brittle member, and the viscous friction, due to the ductile member.  This is not the subject of this paper, but is the topic of our ongoing research into the statistical mechanics of fracture joint friction.

Because of unrelated physics, the same property, ductile fraction, is one of two important order parameters for the linear, isotropic, elastic response of binary mixtures of a structurally competent member (high coordination number) and a structurally less competent member (lower coordination number).  A very important implication of this bicritical model is that the state is only two dimensional.  The expectation, and practical reality (as demonstrated by analysis of well log data) is that the isotropic properties will reduce to a surface in the three dimensional density, compressional velocity, shear velocity (i.e., $\rho, v_p, v_s$) space.  Furthermore, this surface will be orthogonal to the $v_p$-$v_s$ plane.  This remarkable property is captured by the floating grain model \citep{demartini.glinsky.06,gunning.glinsky.07} which has two state variables given by the floating grain fraction, $f_f$, and the compaction state as specified by $1-\mathrm{exp}(-P_e/P_0)$, where $P_e$ is the effective stress and $P_0$ is a reference value of effective stress.  Two phase transitions points at critical values in the radius ratio (${RR}_c=4$) and the fraction of small grains (${VF}_c=0.45$) were demonstrated, as well as two critical scalings of the porosity about a critical point of about 42\% \citep{bryant.etal.09}.

This theory was developed for a binary mixture of brittle spheres of two different sizes.  Recognizing that the large spheres are the structurally competent member and the small spheres are the structurally less competent member, we generalize this theory in Sec. \ref{rock.physics.section}.  The floating grain fraction is replaced by a general geometry parameter, $\xi$, which in the case of shales is shown to be proportional to $f_d$, where $f_d$ is the ductile fraction.  The geometry parameter captures the fabric of the mixture such as the sorting or ductile fraction, while the composition parameter captures the compaction, diagenesis, and/or mineral substitution of the mixture.  An important additional implication of the bicritical model is a fundamental self similarity and the associated scaling relationships\citep{stauffer.aharony.94} of physical quantities such as coordination numbers, capture fractions, and elastic moduli.  It also implies the same critical scaling for both $v_p$ and $v_s$ because they have the same units.  Therefore the surface in $(\rho, v_p, v_s)$ space must be orthogonal to the $v_p$-$v_s$ plane.

We emphasize the serendipity of the fact that the ductile fraction is the coordinate of influence of both the linear elastic response (geophysical) and the nonlinear inelastic response (geomechanical).  For the former, the ductile material is adding density without much structural rigidity, that is elastic moduli.  For the latter, it is increasing the importance of the viscous joint friction.

Given this rock physics model, this paper examines its implication on the geophysical detectability of ductile fraction in Sec. \ref{forward.model.section}.  Several questions have been the subject of much debate within the geophysical community \citep{goodway.10,veire.landro.06,mahmoudian.etal.04,stewart.etal.02,khadeeva.vernik.13,hornby.etal.94,sayers.13,vernik.kachanov.10,sayers.05,guo.etal.13}.  For example, how many stacks should be used in ``prestack'' analysis?  What should those stacks be?  What is the relative value of AVO versus converted wave data analysis?  What is the value of determining density from large angle PP data?  What are the quantities that should be inverted for, relative (reflectivity) versus absolute (impedances)?  Finally, what are the ``attributes'' that best predict reservoir performance?

We present a straight forward analytic theory in Sec. \ref{svd.section} and subsequent analysis that answers all of these questions in Sec. \ref{svd.analysis.section}.  It is a linear singular value decomposition analysis\citep{saleh.debruin.00,causse.etal.07a,causse.etal.07b,varela.etal.09} of the relationship between the two fundamental rock physics parameters ($\zeta$ and $\xi$) and the seismic reflectivities (PP and PS) as functions of angle of incidence, $\theta$.  This analysis is done by assuming a weak scattering (Born) approximation and plane wave assumption (eikonal).  It also orders the SVD using the angle $\theta$.  Distortions caused by angle dependent noise and by angle dependent multiplicative factors are also examined.  The conclusion is that the PP full stack and the PS ``full'' (linear weighted with $\theta$ or offset) stacks are optimal in the estimation of $\zeta$ and $\xi$, respectively.  They are of zeroth and first order in $\theta$, respectively.  Conventional AVO ``gradient'' stacks and large angle PP response conventionally used to estimate density are of higher order in $\theta$ (second and fourth order, respectively).  Angle dependent noise and multiplicative distortion modify the weights of the full stacks and practically lead to the common taper and offset dependent scalars used.  It should be noted that these stacks are average reflectivities or relative quantities.  Linear combinations of these two important stacks (normally just the full PP stack for $\zeta$ and the ``full'' PS stack for $\xi$) are the best ``attributes''.

Although this analysis is expanded to 5th order in the sine function of the maximum angle of incidence, $\sin{\theta_m}$, and the expressions are valid to arbitrary large angle; they have tenuous validity at large angle because of an increasing difficulty in satisfying the eikonal and weak scattering approximations at larger angles.  Another manifestation of this is the inability to renormalize the theory (average at different scales).  To correct this we formally truncate the theory at second order (in the latter part of Sec. \ref{forward.model.section}) and introduce renormalization coefficients that are essentially changes in PP wavelet amplitude, PS wavelet amplitude, and effective incident angle as a function of scale.  This theory is well known to be renormalizable.  A practical implication is that it can be shown to closely match the full wave solution.  It is then shown that a synthetic can be constructed using separate effective wavelets for each of the three stacks (i.e., full PP, ``full'' PS, and AVO PP ``gradient'' stacks).  This allows us to conveniently derive the separate wavelets and renormalization constants by a conventional wavelet derivation process\citep{gunning.glinsky.06}  using the well logs and corresponding measured seismic data -- it allows us to separate the wavelet from the reflectivity analysis.

Finally, the practical detectability, on a synthetic example based on the Marcellus shale play, is shown in Sec. \ref{applications.section}.  There are many factors that can complicate and confound this analysis, such as tuning effects of multiple layers, low SNR in real data, and uncertainty in the rock physics model.  To address these issues on a prototypical example, a principle components analysis and wavelet derivation on real data are done in Sec. \ref{stack.weights.section}.  This includes stack weight profiles, spectral SNR analysis and wavelet profiles.  The uncertainty of the rock physics model is estimated using reasonably large well log database from several unconventional shale plays.  First, the SVD analysis is extended to include the rock physics uncertainty in Sec. \ref{detectability.rock.physics.section} and the detectability of the rock physics parameters, $\zeta$ and $\xi$ is determined.  Second, it is used to construct a layer based model of the Marcellus play with uncertainty (in Sec. \ref{marcellus.model.section}), to forward model the synthetic, and finally to do a layer based Bayesian inversion\citep{gunning.glinsky.04,chen.glinsky.13}  of this model (in Sec. \ref{model.inversion.section}).  Very good sensitivity to the ductile fraction is found in the high TOC (Total Organic Carbon) shale layers.  Significant additional sensitivity is found by using the ``full'' PS data, in addition to the full PP data.

\section{Theory}
\label{theory.section}

\subsection{Rock physics}
\label{rock.physics.section}

We first recognize that we are dealing with a binary mixture of a ductile and a brittle member, where the latter is more structurally competent than the former.  We take inspiration from the floating grain model \citep{demartini.glinsky.06}.  This model is based on two fundamental parameters -- the floating grain fraction parameterized by $\xi=f_f/f_{fc}$ and the compaction parameterized by $\zeta=1-\mathrm{exp}(-P_e/P_0)$, where $f_f$ is the floating grain fraction, $f_{fc}$ is the maximum or critical floating grain fraction, $P_e$ is the effective stress, and $P_0$ is a reference effective stress.  The model respects fluid substitution and leads to local linear correlations of the form
\begin{align}
\label{vp.trend.eqn}
v_p &= A_{vp} + B_{vp} \, \zeta + C_{vp} \, \xi \; \pm \sigma_{vp}, \\
\label{phi.trend.eqn}
\phi &= A_{\phi} + B_{\phi} \,  v_p + C_{\phi} \, \xi \; \pm \sigma_{\phi},  \; \text{and} \\
\label{vs.trend.eqn}
v_s &= A_{vs} + B_{vs} \, v_p \; \pm \sigma_{vs}.
\end{align}
The second relationship can be rewritten two ways, given $\rho_s$ and $\rho_f$ and the definition $\rho \equiv \phi \, \rho_f + (1-\phi) \rho_s$,
\begin{align}
\label{frac.phi.trend.eqn}
\phi &= \phi_c - \frac{\phi_c}{n_{\zeta}} \, \zeta - \frac{\phi_c}{n_\xi} \, \xi, \; \text{and} \\
\label{rho.trend.eqn}
\rho &= A_\rho + B_\rho \, v_p + C_\rho \, \xi \; \pm \sigma_\rho.
\end{align}
The first, Eq. (\ref{frac.phi.trend.eqn}), identifies the two critical exponents, $n_\zeta$ and $n_\xi$, and the critical porosity, $\phi_c$, in the linear expansion, as $\phi/\phi_c \to 0$, of the following expressions for the critical scalings of $\zeta$ and $\xi$, respectively:
\begin{equation}
\zeta \sim \left( \frac{\phi_c - \phi}{\phi_c} \right)^{n_\zeta} \; \text{and} \; \; \xi \sim \left( \frac{\phi_c - \phi}{\phi_c} \right)^{n_\xi}.
\end{equation}
The second, Eq. (\ref{rho.trend.eqn}), is just a convenient expression to compare to log data of shales.

For the rocks studied in \citet{demartini.glinsky.06}, the regressed values are given by $A_{vp} =$ 5000 ft/s, $B_{vp} =$ 6720 ft/s, $C_{vp} =$ 1603 ft/s, $\sigma_{vp} =$ 350 ft/s, $A_\phi =$ 0.592, $B_\phi =-3.14 \times 10^{-5}$ s/ft, $C_\phi =$ -0.0878, $\sigma_\phi =$ 0.0093, $A_{vs} =$ -2900 ft/s, $B_{vs} =$ 0.894, $\sigma_{vs} =$ 226 ft/s, $\phi_c =$ 0.435, $n_\zeta =$ 2.06, $n_\xi =$ 3.11, $P_0 =$ 1290 psi, $f_{fc} =$ 0.09, $A_\rho =$ 1.69 gm/cc, $B_\rho = 5.33 \times 10^{-5}$ (s/ft)(gm/cc), $C_\rho =$ 0.149 gm/cc, and $\sigma_\rho =$ 0.016 gm/cc.  We have assumed $\rho_s =$ 2.7 gm/cc, and $\rho_f =$ 1.0 gm/cc in these relationships.  Note that $\phi_c$ is the expected percolation threshold.  A very important property of this model is the form of the $v_s$ correlation -- it is only a function of $v_p$ and does not involve either $\zeta$ or $\xi$.  This means that the rock physics correlates the $\rho$, $v_p$, and $v_s$ values into a plane that is orthogonal to the $v_p$-$v_s$ plane.  Characteristic values for the rock physics parameters are $\zeta = 0.910 \pm 0.012$ and $\xi = 0.22 \pm 0.33$.
\begin{figure}
\noindent\includegraphics[width=20pc]{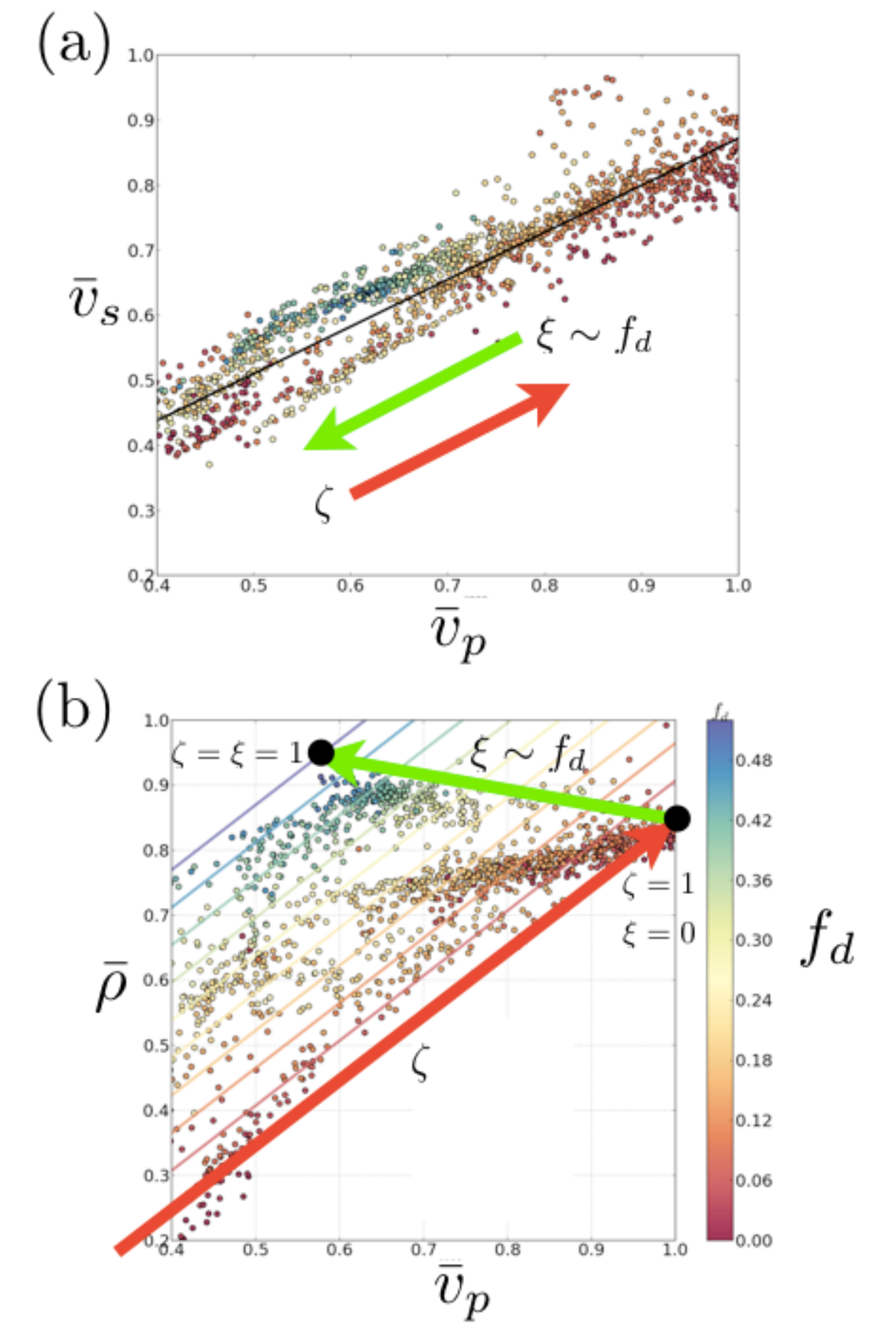}
\caption{\label{fig1} Well log data supporting rock physics model. Points are blocked well log data colored according to the ductile fraction, $f_d$.  Also shown are the directions of increasing $\zeta$ (constant $\xi$) as the red arrow, and increasing $\xi$ (constant $\zeta$) as the green arrow. Values are normalized according to the equation $\bar{x} = (x-x_{min})/(x_{max}-x_{min})$, where $\min{v_p}=$ 8000 ft/s, $\max{v_p}=$ 18000 ft/s, $\min{v_s} =$ 3800 ft/s, $\max{v_s}=$ 11000 ft/s, $\min{\rho}=$ 2.1 gm/cc, $\max{\rho}=$ 2.8 gm/cc. (a) $v_s$-$v_p$ trend in normalized units.   Black line is the fit trend, Eq. (\ref{vs.trend.eqn}).  (b) $\rho$-$v_p$ trend in normalized units.  Trend lines of constant $\xi$, Eq. (\ref{vp.trend.eqn}), are colored according to the value of $f_d=f_{dc} \xi$.  Two reference points are shown as black dots and labeled. }
\end{figure}

When we examine shales from many different wells and plays, we get the results shown in Fig. \ref{fig1}.  It is important to note the strong linear correlation in the $v_p$-$v_s$ plane of Fig. \ref{fig1}a, and the systematic shift in the $\rho$-$v_p$ correlation with the ductile fraction, $f_d$, in Fig. \ref{fig1}b.  Inspired by the floating grain model, we generalize $\xi$ to $f_d/f_{dc}$, where $f_d$ is the ductile fraction and $f_{dc}$ is the maximum or critical ductile fraction.  Ductile fraction is defined as the ratio of the structurally incompetent (ductile) organic matter (TOC) and clay, to the sum of the structurally incompetent plus the structurally competent (brittle) quartz and calcium carbonate.  Given the range of the data, the line in Fig. \ref{fig1}b shows the variation in the $\rho$-$v_p$ as $\xi$ goes from 0 to 1 for $\zeta=1$, and we assume that the minimum value of $v_p$ is 9500 ft/s when $\zeta=\xi=0$.  A regression to this extended rock physics model leads to $A_{vp} =$ 9500 ft/s, $B_{vp} =$ 8500 ft/s, $C_{vp} =$ -4500 ft/s, $\sigma_{vp} =$ 350 ft/s, $A_\phi =$ 0.771, $B_\phi =-3.68 \times 10^{-5}$ s/ft, $C_\phi =$ -0.1916, $\sigma_\phi =$ 0.017, $A_{vs} =$ 1280 ft/s, $B_{vs} =$ 0.48, $\sigma_{vs} =$ 216 ft/s, $\phi_c =$ 0.421, $n_\zeta =$ 1.34, $n_\xi =$ 16.4, $f_{dc} =$ 0.52, $A_\rho =$ 1.435 gm/cc, $B_\rho = 7.0 \times 10^{-5}$ (s/ft)(gm/cc), $C_\rho =$ 0.364 gm/cc, and $\sigma_\rho =$ 0.032 gm/cc.  We have assumed $\rho_s =$ 2.9 gm/cc and $\rho_f=$ 1.0 gm/cc in these relationships.  We note that there was a four-fold decrease in $\sigma_\rho$ by including the $C_\rho$ term in the regression of the datasets.  Note the reasonable value of $\phi_c$.  The self similarity of the rock structure implied by this model is validated by neutron scattering experiments \citep{clarkson.etal.13}.  Characteristic values of the rock physics parameters are $\zeta=0.75 \pm 0.07$ and $\xi=0.70 \pm 0.20$.  Straight forward analysis shows that the capture fraction, as defined by \citet{demartini.glinsky.06}, scales as $n_\xi / (n_\xi-n_\zeta)$, is approximately equal to the reciprocal of this exponent for states away from the critical point, and is the ratio of the ductile coordination number to the brittle coordination number.  This gives a capture fraction of 92\% for this model, and 36\% for the floating grain work of \citet{demartini.glinsky.06}.

We have not explicitly identified the process and therefore the ``activation energy'' in the definition of $\zeta \equiv 1- \mathrm{exp}(-E/E_0)$.  Unlike for the floating grain model, changes in the composition are not limited to compaction (there is probably a very large amount of diagenesis and mineral substitution for shales), and we did not have information on what the controlling variables (i.e., effective stress or temperature) were for each of the well log samples.  Practically, this is not a limitation since we are not trying to estimate the energy, $E$, and that value will be assumed to be a constant for a stratigraphic layer in our analysis.  This, not withstanding, there is a strong possibility that if the compaction and diagenesis are constant for a stratigraphic interval, the composition variable would be diagnostic of the organic matter (TOC) to clay ratio.

The relationship for the shift in the $\rho$ vs. $v_p$ trend, given Eq. (\ref{rho.trend.eqn}), or equivalently the $\phi$ vs. $v_p$ trend, given by Eq. (\ref{phi.trend.eqn}), with clay fraction has also been noted by \citet{han.etal.86} and \citet{pervukhina.etal.13} in laboratory core data.

\begin{widetext}
\subsection{Geophysical forward model}
\label{forward.model.section}

We now move to developing an understanding of both the P-to-P, $R_{PP}$, and the P-to-S, $R_{PS}$, reflection response for an isotropic medium.  We start by assuming weak scattering and make both the further assumptions of small contrast (that is, $\Delta \rho/\rho, \Delta v_p/v_p, \, \text{and} \,\Delta v_s/v_s \ll 1$) and plane waves (eikonal approximation).  The latter is a rather complicated assumption on both the frequency and angle of incidence, $\theta$.  We shall return to this later in this section.  The expressions \citep{shaw.sen.04} for the reflection response will be linear in the contrasts due to the first approximation, with coefficients that are functions of the angle of incidence, $\theta$, and the ratio of the velocities, $r_{sp} \equiv v_s/v_p$,
\begin{align}
R_{PP}&=\frac{1}{2} \left( \frac{\Delta \rho}{\rho}+\frac{\Delta v_p}{v_p} \right) + \left( -2 \, r_{sp}^2 \frac{\Delta \rho}{\rho}+\frac{1}{2}\frac{\Delta v_p}{v_p}-4 \, r_{sp}^2\frac{\Delta v_s}{v_s} \right) \sin^2{\theta} +\frac{1}{2}\frac{\Delta v_p}{v_p} \sin^2{\theta} \tan^2{\theta},
\\
R_{PS}&=-\frac{\sin{\theta}}{\cos{\theta_{PS}}} \left[ \frac{1}{2} \frac{\Delta \rho}{\rho} +\left(\frac{\Delta \rho}{\rho}+2\frac{\Delta v_s}{v_s} \right)   \left( r_{sp} \cos{\theta} \cos{\theta_{PS}} - r_{sp}^2 \sin^2{\theta} \right) \right],
\end{align}
where $\theta_{PS}$ is the reflected angle of the S wave.  Making use of Snell's law,
\begin{equation}
\frac{\sin{\theta_{PS}}}{v_s}=\frac{\sin{\theta}}{v_p},
\end{equation}
some basic trigonometric identities and combining terms of common order in $\sin{\theta}$, the reflectivities can be written as
\begin{align}
\label{eqn.pp.sin}
R_{PP}&=\frac{1}{2} \left( \frac{\Delta \rho}{\rho}+\frac{\Delta v_p}{v_p} \right) + \left( -2 \, r_{sp}^2 \frac{\Delta \rho}{\rho}+\frac{1}{2}\frac{\Delta v_p}{v_p}-4 \, r_{sp}^2\frac{\Delta v_s}{v_s} \right) \sin^2{\theta} +\frac{1}{2}\frac{\Delta v_p}{v_p} \frac{\sin^4{\theta} }{\cos^2{\theta}},
\\
\label{eqn.ps.sin}
\begin{split}
R_{PS}&=\left[ \left(-\frac{1}{2}-r_{sp}\right)\frac{\Delta \rho}{\rho} -2 \, r_{sp} \frac{\Delta v_s}{v_s} \right] \frac{\sin{\theta}}{\sqrt{1-(r_{sp} \sin{\theta})^2}} \\&+\left[\frac{r_{sp}}{2}(1+r_{sp})^2 \frac{\Delta \rho}{\rho} + r_{sp} (1+r_{sp})^2\frac{\Delta v_s}{v_s} \right] \frac{\sin^3{\theta}}{\sqrt{1-(r_{sp} \sin{\theta})^2}} \\ &+ \left(r_{sp}\frac{\Delta \rho}{\rho}+2 r_{sp}\frac{\Delta v_s}{v_s} \right) \left[ \frac{1-\cos{\theta}\sqrt{1-(r_{sp} \sin{\theta})^2}}{\sin^2{\theta}}-\frac{1}{2}(1+r_{sp}^2) \right]  \frac{\sin^3{\theta}}{\sqrt{1-(r_{sp} \sin{\theta})^2}} .
\end{split}
\end{align}
Expanding to the 4th order in $\theta$ leads to the expressions
\begin{align}
\label{eqn.pp.theta}
R_{PP}&=\frac{1}{2} \left( \frac{\Delta \rho}{\rho}+\frac{\Delta v_p}{v_p} \right) + \left( -2 \, r_{sp}^2 \frac{\Delta \rho}{\rho}+\frac{1}{2}\frac{\Delta v_p}{v_p}-4 \, r_{sp}^2\frac{\Delta v_s}{v_s} \right) \theta^2 + \left( \frac{2}{3}r_{sp}^2 \frac{\Delta \rho}{\rho}+\frac{1}{3}\frac{\Delta v_p}{v_p}+\frac{4}{3}r_{sp}^2\frac{\Delta v_s}{v_s} \right) \theta^4 + \mathcal{O}(\theta^6),
\\
\label{eqn.ps.theta}
R_{PS}&=\left[ \left(-\frac{1}{2}-r_{sp}\right)\frac{\Delta \rho}{\rho} -2 \, r_{sp} \frac{\Delta v_s}{v_s} \right] \theta + \left[\left(\frac{1}{12}+\frac{2}{3}r_{sp}+\frac{3}{4}r_{sp}^2\right)\frac{\Delta \rho}{\rho} \right. \left. + \left(\frac{4}{3}r_{sp}+2 \, r_{sp}^2 \right)\frac{\Delta v_s}{v_s} \right] \theta^3 + \mathcal{O}(\theta^5).
\end{align}
We have been careful to write these expressions in a bilinear form in terms of the small contrast (i.e., $\Delta \rho / \rho, \Delta v_p /  v_p, \Delta v_s / v_s$) and the angle of incidence (i.e., $\sin^n{\theta}$ or $\theta$).  This will facilitate the SVD analysis of the next section.  Note that the coefficients of this bilinear transformation are only functions of the dimensionless parameter, $r_{sp}$.
\end{widetext}

Before we continue our analysis, we take a closer look at the plane wave (or eikonal) portion of the weak scattering (or Born) approximation.  This is a quite non-trivial assumption that puts an upper limit on the validity of the $\theta$, given by the condition that the dimensionless scale of the perturbation
\begin{equation}
\frac{\lambda}{T \cos{\theta}} \equiv s \ll 1,
\end{equation}
where $\lambda$ is the wavelength of the wave and $T$ is the scale of the gradient or the thickness of the layer.  The problem is that this can never be satisfied because there is no well defined scale for the medium.   The question now becomes:  how does the expression for $R_{PP}$ and $R_{PS}$ (which we now call
\begin{equation}
\label{def.r.eqn}
R \equiv (R_{PP}; R_{PS})
\end{equation}
collectively), given in the expansions of Eq. (\ref{eqn.pp.theta}) and Eq. (\ref{eqn.ps.theta}), average as a function of dimensionless scale, $s$?  We now evoke well known theoretical physics concepts of renormalization\citep{maggiore.10}, to recognize that we need to expand in scale about the ``ground state'' harmonic oscillator.  We introduce three running coupling constants $a_0(s)$, $a_1(s)$, and $a_2(s)$; and define the coefficients of the reflectivity, ordered by $\theta^n$ as
\begin{align}
A_0(\Delta c) &= \frac{1}{2} \left( \frac{\Delta \rho}{\rho}+\frac{\Delta v_p}{v_p} \right),
\\
A_1(\Delta c) &= -\left(\frac{1}{2}+r_{sp}\right)\frac{\Delta \rho}{\rho} -2 \, r_{sp} \frac{\Delta v_s}{v_s},
\\
A_2(\Delta c) &= -2 \, r_{sp}^2 \frac{\Delta \rho}{\rho}+\frac{1}{2}\frac{\Delta v_p}{v_p}-4 \, r_{sp}^2\frac{\Delta v_s}{v_s},
\end{align}
where the small contrasts $\Delta c \equiv (\Delta \rho / \rho, \Delta v_p / v_p, \Delta v_s / v_s)$ are taken at the same reference scale, $s$.  The reflectivity at a scale, $s$, can now be written as
\begin{equation}
R=[a_0 A_0 + a_2 A_2 \theta^2; a_1 A_1 \theta].
\end{equation}
Another way of looking at this is a redefinition of incidence angle, $\overline{\theta} \equiv \theta \sqrt{a_2/a_0}$, and reflection coefficient, $\overline{R} \equiv [R_{PP}/a_0; R_{PS}/(a_1 \sqrt{a_0/a_2})]$ so that
\begin{equation}
\overline{R}=[A_0+A_2 \overline{\theta}^2; A_1 \overline{\theta}].
\end{equation}
The relationship between these expressions is just that of dressed to undressed fields.  In the case that there is a well defined scale and $\theta$ is small enough, $a_0=a_1=a_2=1$.  Otherwise, one must calculate the running coupling constants for the scale of interest using a characteristic well log of the isotropic elastic properties and a forward wave solution with a wavelet of scale, $\lambda$.

Recognizing that we will be truncating the expansion at the second order in $\theta$, we now develop a convenient approximation to the forward model of a spike convolution
\begin{equation}
R(\theta ; t) = \sum_k{R(\theta ; \Delta c_k) \; W(\theta ; t-t_k)},
\end{equation}
where $W(\theta;t)$ is a given angle dependent wavelet and the summation is over the $\{k\}$ contrasts or interfaces.  The problem with this expression is the $\theta$ dependance of the wavelet.  We would like to eliminate it, and replace it by average wavelets.  To this end, we now decompose $R(\theta; \Delta c)$ according to its $\theta$ dependance.  Given the simple form of Eq. (\ref{eqn.pp.theta}) and Eq. (\ref{eqn.ps.theta}) it would have three singular values $\lambda_i$ and singular vectors $\xi_i(\theta)$.  For a more general expansion as given in Eq. (\ref{eqn.pp.sin}) and Eq. (\ref{eqn.ps.sin}), it would have more singular values, $\lambda_i \sim \mathcal{O}(\theta_m^i)$, where $\theta_m$ is the maximum angle of incidence.  This structure will be analyzed in more detail in the next section.  For now we just project $R(\theta; t)$ onto this basis
\begin{equation}
R_i(t) \equiv \int{\xi_i(\theta) \; R(\theta; t) \; d\theta},
\end{equation}
and define
\begin{align}
W_i(t) &\equiv \int{\xi_i(\theta) \; W(\theta; t) \; d\theta},
\\
\Delta W_i(\theta; t) &\equiv W(\theta; t) - W_i(t),
\\
R_i(\Delta c) &\equiv \int{\xi_i(\theta) \; R(\theta; \Delta c) \; d\theta},
\\
\Delta R_i(\theta; \Delta c) &\equiv R(\theta; \Delta c) - R_i(\Delta c).
\end{align}
Remember that to second order
\begin{equation}
R(\theta; \Delta c) = [a_0 A_0(\Delta c) + a_2 A_2(\Delta c) \theta^2; a_1 A_1(\Delta c) \theta] + \mathcal{O}(\theta^3).
\end{equation}
Recognizing that
\begin{equation}
\int{\xi_i(\theta) \; \Delta W_i(\theta; t) \; d\theta} = \int{\xi_i(\theta) \; \Delta R_i(\theta; t) \; d\theta}  = 0
\end{equation}
and that $\Delta W_i$ and $\Delta R_i$ are of second order in $\theta^2$, we find that
\begin{align}
\begin{split}
R_i(t) &= \sum_k{\int{d\theta \; \xi_i(\theta) \left[ R_i(\Delta c_k)+\Delta R_i(\theta;\Delta c_k) \right]}} \\ & \quad\quad\quad\quad\quad \left[ W_i(t-t_k)+\Delta W_i(\theta; t-t_k) \right] 
\end{split}
\\
\begin{split}
&= \sum_k{\biggl[ R_i(\Delta c_k) \; W_i(t-t_k) \biggr. } \\ & \quad \left. + \int{d\theta \; \xi_i(\theta) \; \Delta R_i(\theta; \Delta c_k) \; \Delta W_i(\theta; t-t_k)} \right]
\end{split}
\\
&= \sum_k{R_i(\Delta c_k) \; W_i(t-t_k)} + \mathcal{O}(\theta_m^4)
\end{align}
This is an extremely convenient result.  What it allows us to do is calculate an effective wavelet, $W_i(t)$, for each weighted stack, $R_i(t)$.  We can then form a simple spike convolution forward model using the singular vectors of the reflectivity, $R_i(\Delta c)$ for each stack.  The order of $R_i(\Delta c)$ will be $\theta_m^i$.  We will therefore be able to use this separation of $R$ and $W$ up to third order in $R_i(\Delta c)$.

\subsection{Singular value decomposition theory}
\label{svd.section}

We now move onto understanding the relationships between the basic rock physics parameters we wish to know, $\xi$ and $\zeta$, and the geophysical measurements.  We do this by establishing a sequence of linear transformations, then examining the important singular value decompositions (SVDs) of that compound transformation.  The singular values will give an understanding of detectability of the singular vectors (that is, the required SNR).  The singular vectors will tell us what views of the measurement to use and how they are related to the rock physics.

We start by writing the expression for the measured reflectivity in following linear form
\begin{equation}
R_m = D M_\theta ( M_A ( M_{RP} \Delta r + \varepsilon_r ) + \varepsilon_A ) + \varepsilon_m,
\end{equation}
where $R_m$ is the measured value of $R$, $D$ is a linear distortion of the measurement of $R$, $M_\theta$ is the angle matrix, $M_A$ is the geophysical reflection matrix, $M_{RP}$ is the rock physics matrix, $\Delta r$ is the change in the rock physics parameters, $\varepsilon_r$ is the error vector in the rock physics relationships, $\varepsilon_A$ is the error vector in the geophysical forward model, and $\varepsilon_m$ is the error vector in the measurement of $R$.  Now expand this expression,
\begin{align}
\begin{split}
R_m &= D M_\theta M_A M_{RP} \Delta r + D ( M_\theta M_A \varepsilon_r + M_\theta \varepsilon_A ) + \varepsilon_m 
\end{split}
\\
\begin{split}
&= R_0 + \varepsilon_c
\end{split}
\end{align}
using the definition of the most likely reflection coefficients
\begin{equation}
\label{RO.eqn}
R_0 \equiv D M_\theta M_A M_{RP} \Delta r
\end{equation}
and the combined error in the estimate of the reflection coefficients
\begin{equation}
\varepsilon_c \equiv D ( M_\theta M_A \varepsilon_r + M_\theta \varepsilon_A ) + \varepsilon_m.
\end{equation}
Assume that the expected values of the fundamental errors of $\varepsilon_r$, $\varepsilon_A$, and $\varepsilon_m$ are 0;  the covariances are given by $\Sigma_r$, $\Sigma_A$, and $\Sigma_m$ respectively; and that $\varepsilon_r$, $\varepsilon_A$, and $\varepsilon_m$ are independent and normally distributed.  It follows that expected value $\varepsilon_c$ is 0, and the covariance is given by 
\begin{equation}
\Sigma_c = \Sigma_m + ( D M_\theta ) \Sigma_A ( D M_\theta )^T +  ( D M_\theta M_A ) \Sigma_r ( D M_\theta M_A )^T 
\end{equation}
In other words, the measurement of the reflection coefficients is distributed according to a multivariant normal distribution, $\text{MVN}(R_0, \Sigma_c)$, with a probability density given by
\begin{equation}
\label{PRm.eqn}
P(R_m) \sim \text{exp}\left\{ -\frac{1}{2} (R_m-R_0)^T \Sigma_c^{-1} (R_m-R_0) \right\}
\end{equation}

\begin{widetext}
Before we continue with understanding the linear structure of this distribution we need to examine the structure of the expression for $R_0$ given in Eq. (\ref{RO.eqn}), and the rock physics covariance matrix, $\Sigma_r$.  First of all the expression for $R_0$ contains the product of matrices where
\begin{equation}
\Delta r \equiv \begin{pmatrix} d\zeta \\ d\xi \end{pmatrix}, \quad \Delta c \equiv \begin{pmatrix} \frac{\Delta \rho}{\rho} \\ \frac{\Delta v_p}{v_p} \\ \frac{\Delta v_s}{v_s} \end{pmatrix},
\quad A \equiv \begin{pmatrix} A_0 \\ A_1 \\ A_2 \\ \vdots \end{pmatrix}, \quad R = \begin{pmatrix} R_{PP}(\theta=0) \\ \vdots \\ R_{PP}(\theta_m) \\ R_{PS}(\theta=0) \\ \vdots \\ R_{PS}(\theta_m) \end{pmatrix},
\end{equation}
\begin{equation}
\label{matrix.eqn}
R = M_\theta \, A, \quad A = M_A \, \Delta c, \quad  \Delta c = M_{RP} \, \Delta r, \; \text{and} \quad
M_{RP} =\begin{pmatrix} \frac{B_\rho \, B_{vp}}{\rho} & \frac{C_{vp}+C_\rho}{\rho} \\ \frac{B_{vp}}{v_p} & \frac{C_{vp}}{v_p} \\ \left(\frac{B_{vs}}{r_{sp}}\right) \frac{B_{vp}}{v_p} & \left(\frac{B_{vs}}{r_{sp}}\right) \frac{C_{vp}}{v_p}  \end{pmatrix} .
\end{equation}
As we have noted in the last section, all of the important physics is contained in the renormalized, 2nd order  in $\theta_m$, (3 term) expressions.  For this case,
\begin{equation}
\label{three.term.eqn}
M_\theta = \begin{pmatrix} 1&0&0 \\ 1&0& (\Delta \theta)^2 \\ 1&0& (2 \Delta \theta)^2 \\ \vdots & \vdots & \vdots \\ 1&0&[(N-2) \Delta \theta]^2 \\ 1&0& \theta_m^2 \\ 0&0&0 \\ 0& \Delta \theta & 0 \\ 0& 2 \Delta \theta & 0 \\ \vdots & \vdots & \vdots \\ 0 & (N-2) \Delta \theta & 0 \\ 0 & \theta_m & 0 \end{pmatrix}, \quad\quad
M_A = \begin{pmatrix} \frac{1}{2} & \frac{1}{2} & 0 \\ -\frac{1}{2} - r_{sp} & 0 & -2 r_{sp} \\ -2r_{sp}^2 & \frac{1}{2} & -4 r_{sp}^2 \end{pmatrix},
\end{equation}
where $\Delta \theta \equiv \theta_m / (N-1)$.  It can be extended to 4th order (5 term) in $\theta_m$ to give
\begin{equation}
\label{five.term.eqn}
M_\theta = \begin{pmatrix} 1&0&0&0&0 \\ 1&0& (\Delta \theta)^2&0& (\Delta \theta)^4 \\ 1&0& (2 \Delta \theta)^2&0& (2 \Delta \theta)^4 \\ \vdots & \vdots & \vdots & \vdots & \vdots \\ 1&0&[(N-2) \Delta \theta]^2&0&[(N-2) \Delta \theta]^4 \\ 1&0& \theta_m^2&0& \theta_m^4 \\ 0&0&0&0&0 \\ 0& \Delta \theta & 0& (\Delta \theta)^3 & 0 \\ 0& 2 \Delta \theta & 0& (2 \Delta \theta)^3 & 0 \\ \vdots & \vdots & \vdots  & \vdots & \vdots \\ 0 & (N-2) \Delta \theta & 0 & [(N-2) \Delta \theta]^3 & 0 \\ 0 & \theta_m & 0 & \theta_m^3 & 0 \end{pmatrix},
M_A = \begin{pmatrix} \frac{1}{2} & \frac{1}{2} & 0 \\ -\frac{1}{2} - r_{sp} & 0 & -2 r_{sp} \\ -2r_{sp}^2 & \frac{1}{2} & -4 r_{sp}^2 \\ \frac{1}{12}+\frac{2}{3} r_{sp} + \frac{3}{4} r_{sp}^2 & 0 & \frac{4}{3} r_{sp}+2 r_{sp}^2 \\ \frac{2}{3} r_{sp}^2 & \frac{1}{3} & \frac{4}{3} r_{sp}^2 \end{pmatrix}.
\end{equation}
We can also give a large $\theta_m$ version extended to 5th order (6 term) in $\sin{\theta_m}$
\begin{equation}
\label{six.term.eqn}
M_\theta^T = \left( \begin{array}{c|c} 1&0 \\ \hline 0 & \frac{\sin{\theta}}{\sqrt{1-(r_{sp} \sin{\theta})^2}} \\ \hline  \sin^2{\theta} & 0 \\ \hline 0 & \frac{\sin^3{\theta}}{\sqrt{1-(r_{sp} \sin{\theta})^2}} \\ \hline   \frac{\sin^4{\theta}}{\cos^2{\theta}} & 0 \\ \hline 0 & \left[ \frac{1-\cos{\theta}\sqrt{1-(r_{sp} \sin{\theta})^2}}{\sin^2{\theta}}-\frac{1}{2}(1+r_{sp}^2) \right]  \frac{\sin^3{\theta}}{\sqrt{1-(r_{sp} \sin{\theta})^2}}  \end{array} \right),
M_A = \begin{pmatrix} \frac{1}{2} & \frac{1}{2} & 0 \\ -\frac{1}{2} - r_{sp} & 0 & -2 r_{sp} \\ -2r_{sp}^2 & \frac{1}{2} & -4 r_{sp}^2 \\ \frac{r_{sp}}{2}(1+r_{sp})^2 & 0 & r_{sp}(1+r_{sp})^2  \\ 0 & \frac{1}{2} & 0 \\ r_{sp} & 0 & 2 r_{sp} \end{pmatrix}.
\end{equation}
Each block of the $M_\theta^T$ matrix is an $1 \times N$ matrix with an element for each discrete $\theta$ between 0 and $\theta_m$.

We do note the degeneracy in the $M_A$ matrix for $r_{sp}=0$ and $1/2$.  This only reduces the rank of $M_A$ to 2 at $r_{sp}=1/2$.  Since $\Delta r$ is only of dimension 2, there is no loss of sensitivity of $R$ to $\Delta r$.

Using Eqs. (\ref{vp.trend.eqn}), (\ref{vs.trend.eqn}) and (\ref{rho.trend.eqn}), the form of the rock physics covariance can be shown to be
\begin{equation}
\frac{\Sigma_r}{2} = \begin{pmatrix} \frac{\sigma^2_\rho+B^2_\rho \sigma^2_{vp}}{\rho^2}&\frac{B_\rho}{\rho v_\rho} \sigma^2_{vp}&\frac{B_\rho B_{vs}}{\rho v_s} \sigma^2_{vp} \\ \frac{B_\rho}{ \rho v_p} \sigma^2_{vp}&\frac{\sigma^2_{vp}}{v^2_p}&\frac{B_{vs}}{v_p v_s} \sigma^2_{vp} \\ \frac{B_\rho B_{vs}}{\rho v_s} \sigma^2_{vp}&\frac{B_{vs}}{v_p v_s} \sigma^2_{vp}&\frac{\sigma^2_{vs} + B^2_{vs} \sigma^2_{vp}}{v^2_s} \end{pmatrix}.
\end{equation}
\end{widetext}

With these definitions now in hand, we return to the form of the distribution for $R_m$ given in Eq. (\ref{PRm.eqn}).  Since $\Sigma_c$ is positive definite, it can be written as
\begin{equation}
\Sigma_c^{-1} = W_d^T W_d
\end{equation}
We make two singular value decompositions (SVDs) such that
\begin{equation}
W_d D M_\theta = U_1 \Sigma_1 V_1^T 
\end{equation}
and
\begin{equation}
\overline{\Sigma}_1 V_1^T M_A M_{RP} = U_2 \Sigma_2 V_2^T.
\end{equation}
We define $\overline{\Sigma}_1$ and $\overline{\Sigma}_2$ as the square diagnal matrices formed by  dropping the zero rows of $\Sigma_1$ and $\Sigma_2$, respectively.  We also define $\overline{U}_1$ and $\overline{U}_2$ by dropping the corresponding columns of $U_1$ and $U_2$, respectively.

First of all, write the distribution as 
\begin{align}
\begin{split}
P(R_m) &\sim \text{exp} \left\{ -\frac{1}{2} (W_d R_m-W_d R_0)^T (W_d R_m-W_d R_0) \right\}
\end{split}
\\
\begin{split}
&\sim \text{exp} \left\{ -\frac{1}{2} \chi^T \chi \right\}
\end{split}
\end{align}
where
\begin{equation}
\chi \equiv W_d R_m - W_d D M_\theta M_A M_{RP} \Delta r .
\end{equation}
Now make the change of coordinates such that 
\begin{equation}
\chi^* \equiv \overline{U}_2^T \overline{U}_1^T \chi .
\end{equation}
Using these definitions, it can be shown that
\begin{equation}
\chi^T \chi = ( \chi^* )^T \chi^* + H
\end{equation}
where $H$ is not a function of $\Delta r$ (thus, it does not affect the likelihood function of $\Delta r$) and
\begin{align}
\begin{split}
\chi^* &= \overline{U}_2^T ( \overline{U}_1^T W_d) R_m - \overline{\Sigma}_2 V^T_2 \Delta r
\end{split}
\\
\begin{split}
&= (\overline{\Sigma}_2 V_2^T) [\Delta r_0 - \Delta r ] ,
\end{split}
\end{align}
where we define
\begin{equation}
\Delta r_0 \equiv (\overline{\Sigma}_2 V_2^T )^{-1} \overline{U}_2^T ( \overline{U}_1^T W_d ) R_m
\end{equation}
and let
\begin{equation}
\Sigma_{\Delta r}^{-1} \equiv (\overline{\Sigma}_2 V_2^T )^T (\overline{\Sigma}_2 V_2^T ) .
\end{equation}
Given that $R_m$ is the observed forward modeled reflection response of rock properties $r_1$ over $r_0$, such that $\Delta r_0 = r_1-r_0$ and $\Delta r = r - r_0$, the probability of $r$ can be written as the multivariate normal distribution, $\text{MVN}(\Delta r_0, \Sigma_{\Delta r})$, with a probability of $r$ given by
\begin{equation}
\label{Pr.eqn}
P(r) \sim \text{exp} \left\{ -\frac{1}{2} (r-r_1) \Sigma_{\Delta r}^{-1} (r-r_1) \right\} .
\end{equation}

Let us now make some practical identifications.  First, recognize that $\overline{U}_1^T W_d$ transforms $R_m$ into $m$ ``stacks'' where $m$ is the dimension of $A$ matrix (either 3, 5 or 6, for Eq. (\ref{three.term.eqn}), (\ref{five.term.eqn}) or (\ref{six.term.eqn}), respectively).  We will denote these stacks as $R_i$ so
\begin{equation}
\tilde{R} \equiv \begin{pmatrix} R_0 \\ R_1 \\ \vdots \\ R_{m-1} \end{pmatrix},  \text{and} \; \overline{\Sigma}_1 = \begin{pmatrix} \lambda_0 & 0 & 0 & 0 \\ 0& \lambda_1 & 0 &0 \\ 0 & 0 & \ddots & 0 \\ 0 & 0 & 0 & \lambda_{m-1} \end{pmatrix}.
\end{equation}
The signal-to-noise level (SNR) of the stack, $R_i$, is defined as $20 \log_{10}{\lambda_i}$ and $\lambda_i \sim \theta_m^i$.  $V_2^T$ is a $2 \times 2$ matrix that rotates $\Delta r$ so that they are orthogonal, $\Delta \tilde{r} = V_2^T \Delta r$.  Then the $m$ stacks $\tilde{R}$ are projected by $\overline{U}_2^T$ (a $2 \times m$ matrix) onto the two orthogonal rock physics directions.  The two singular values given by the diagonal matrix $\overline{\Sigma}_2$ give the uncertainty of the estimates of the rock physics parameters along the two orthogonal directions in the rock physics space, $\Delta \tilde{r}$, defined by $V_2^T$.  One can directly form the two optimal stacks for estimation of the two orthogonal rock physics parameters, $\tilde{\zeta}$ and $\tilde{\xi}$, by $\overline{U}_2^T \overline{U}_1^T W_d$.

Many of the current inversion schemes invert for various moduli and other elastic parameters such as densities and Poisson ratios.  There have been historical debates on which of these combinations are best to estimate the fundamental rock physics parameters that continue to this day.  It is our view that this is an irrelevant debate.  The relevant question is what are the orthogonal stacks of the data covariance matrix with positive SNR and how are they related to the orthogonal coordinates of the rock physics.  Not withstanding this point, there is something to be learned from examining the linear mapping of the rock physics to contrasts in these traditional variables and the SVD of that transformation.

We start this analysis with the definition of a reasonably representative set of traditional parameters which consists of the shear modulus,
\begin{equation*}
G \equiv \rho \, v_s^2,
\end{equation*}
the bulk modulus,
\begin{equation*}
K \equiv \rho \, v_p^2 - \frac{4}{3} G,
\end{equation*}
the Youngs modulus,
\begin{equation*}
E \equiv \frac{9 K G}{3 K + G},
\end{equation*}
the Poisson ratio,
\begin{equation*}
\nu \equiv \frac{3 K - 2 G}{2(3 K + G)},
\end{equation*}
the $v_p$ to $v_s$ ratio,
\begin{equation*}
r_{ps} \equiv v_p / v_s,
\end{equation*}
and the density, $\rho$.  We linearize the relationship between these variable and $\Delta c$ so that 
\begin{equation}
\label{rt.eqn}
\Delta r_T = M_T \Delta c, 
\end{equation}
where
\begin{equation}
\Delta r_T \equiv \begin{pmatrix} \frac{\Delta K}{K} \\ \frac{\Delta G}{G} \\ \frac{\Delta E}{E} \\ \frac{\Delta r_{ps}}{r_{ps}} \\ r_{ps}^2 \, \Delta \nu \\ \frac{\Delta \rho}{\rho} \end{pmatrix}, \text{and}
\end{equation}
\begin{equation}
M_T =
\left(
\begin{array}{ccc}
 1 & -\frac{6}{4 r_{sp}^2-3} & \frac{8 r_{sp}^2}{4 r_{sp}^2-3} \\
 1 & 0 & 2 \\
 1 & -\frac{2 \left(2 r_{sp}^2-3\right) \left(2 r_{sp}^2-1\right)}{(r_{sp}-1) (r_{sp}+1) \left(4 r_{sp}^2-3\right)} & \frac{2 \left(8 r_{sp}^4-15 r_{sp}^2+6\right)}{(r_{sp}-1) (r_{sp}+1) \left(4 r_{sp}^2-3\right)} \\
 0 & 1 & -1 \\
 0 & \frac{1}{(r_{sp}-1)^2 (r_{sp}+1)^2} & -\frac{1}{(r_{sp}-1)^2 (r_{sp}+1)^2} \\
 1 & 0 & 0 \\
\end{array}
\right).
\end{equation}
This linear relationship is singular for $r_{sp}=1$ and $\sqrt{3/4}$.  It is constructed to have a well defined limit at $r_{sp}=0$ of
\begin{equation}
M_T =
\begin{pmatrix}  1&2&0 \\ 1&0&2 \\ 1&-2&4 \\ 0&-1&1 \\ 0&-1&1 \\ 1&0&0 \end{pmatrix},
\end{equation}
which shows that the moduli (bulk, shear, and Youngs) are mixtures of the density and the velocities, the Poisson ratio and the $v_p$ to $v_s$ ratio are both similar quantities showing correlation in $v_p$ to $v_s$, and the density is modestly perpendicular to the moduli.  These facts will be useful in understanding the results to be shown in Fig. \ref{fig9} in Sec. \ref{svd.analysis.section}.

Using the Eq. (\ref{rt.eqn}) and Eq. (\ref{matrix.eqn}), we write
\begin{equation}
\Delta r_T = M_T M_{RP} \Delta r.
\end{equation}
Now make the SVD, so that $M_T M_{RP} = U_T \Sigma_T V_T^T$.  The $V_T=V_2$ that we found before, so that we write
\begin{equation}
\overline{U}_T^T \Delta r_T = \overline{\Sigma}_T V_2^T \Delta r= \overline{\Sigma}_T  \Delta \tilde{r}.
\end{equation}
The interesting part of this SVD is $\overline{U}_T^T$ which is a $2 \times 6$ matrix which projects the traditional rock physics contrasts onto two orthogonal rock physics directions.

\section{Applications}
\label{applications.section}

\subsection{Singular value decomposition analysis}
\label{svd.analysis.section}

This is still abstract at this point.  Let us substitute in the rock physics of the shales given in the latter part of Sec. \ref{rock.physics.section}.  For now we set the multiplicative distortion, $D$, to the identity matrix and the data covariance, $\Sigma_m$, to a diagonal constant of 1.  We shall return to this later in this section.  Also set the rock physics covariance, $\Sigma_r$, to zero along with the covariance of the forward model, $\Sigma_A$.  We shall return to the implications of rock physics uncertainty on the detectability of ductile fraction in Sec. \ref{detectability.rock.physics.section}.  The matrix $W_d$ will therefore be the identity matrix.  We set the rock physics composition to $\zeta=0.79$ and the geometry to $\xi=0.5$.  This gives a density of $\rho=2.59 \, \text{gm/cc}$, compressional velocity of $v_p= 14000 \, \text{ft/s}$, a shear velocity of $v_s= 8000 \, \text{ft/s}$, a $v_p$ to $v_s$ ratio of $r_{ps}= 1.75$, a Poisson ratio of $\nu= 0.26$, and a porosity of $\phi= 16\%$.  

\begin{figure}
\noindent\includegraphics[width=20pc]{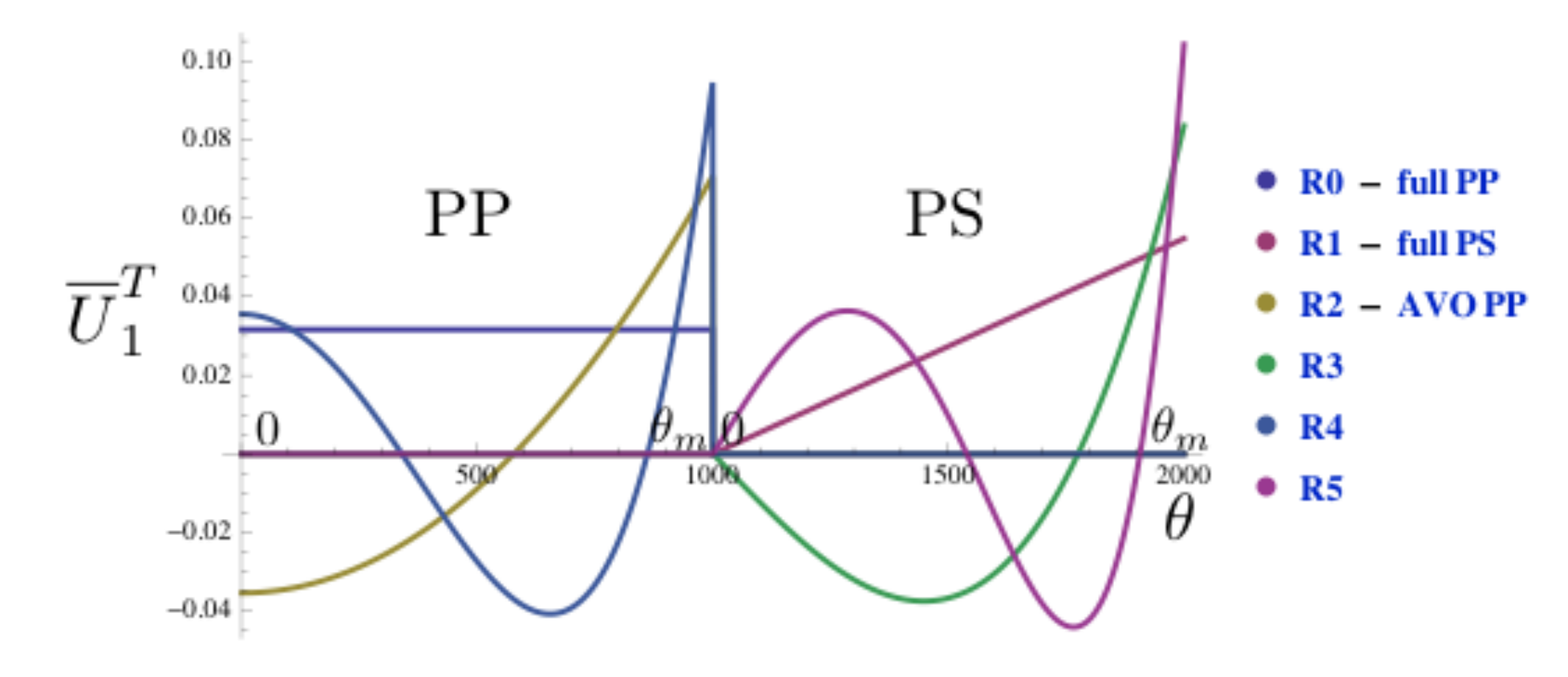}
\caption{\label{fig2} Stack weights, $\overline{U}_1^T$, as a function of incidence angle, $\theta$.  First set is for PP data, followed by the weights for PS data.}
\end{figure}

For a small maximum angle of $\theta_m= 0.5^\circ$, we get the stack weights, $\overline{U}_1^T$, shown in Fig. \ref{fig2}.  We have shown the results for the 6 term $A$ vector, but the other two are just truncated versions of this result.  It should be noted that this result is independent of the rock physics, $M_{RP}$, and the relationship between the rock physics and the $A\text{'s}$, $M_A$.  In the order of decreasing singular value, or SNR, we have $R_0$ the full PP stack, $R_1$ the ``full'' PS stack (in quotes because it is really linearly weighted with $\theta$), then $R_2$ the AVO PP gradient stack (weighted by $\theta^2$ so that it is the far offsets minus the near offsets).  The series continues on with progressively higher $\theta$ order weightings of the stacks in an alternating order between the PP and the PS data.  The next figure (Fig. \ref{fig3}), shows the dependance of the singular values on $\theta_m$.  Note that they scale as $\lambda_i \sim \theta_m^i$ as expected.  Continuing with the analysis, we show the rotation of $\Delta r$ onto an orthogonal system $\Delta \tilde{r}$ in Fig. \ref{fig4}.  Note that $\tilde{\zeta}$ is mainly the composition variable $\zeta$ and the $\tilde{\xi}$ variable is mainly the geometry variable $\xi$.  Figure \ref{fig5} shows the $\overline{U}_2^T$ transformation of the stacks, $\tilde{R}$, onto the rock physics variables, $\Delta \tilde{r}$.  Note that the full PP stack is the main contribution to the determination of the composition variable, $\tilde{\zeta}$, and the ``full'' PS stack is the main contribution to the determination of the geometry variable, $\tilde{\xi}$.  The AVO PP gradient stack is of minor contribution to either, but it is more aligned with $\tilde{\xi}$ and orthogonal to $\tilde{\zeta}$.  The 4th order PP, $R_4$, is totally negligible.
\begin{figure}
\noindent\includegraphics[width=20pc]{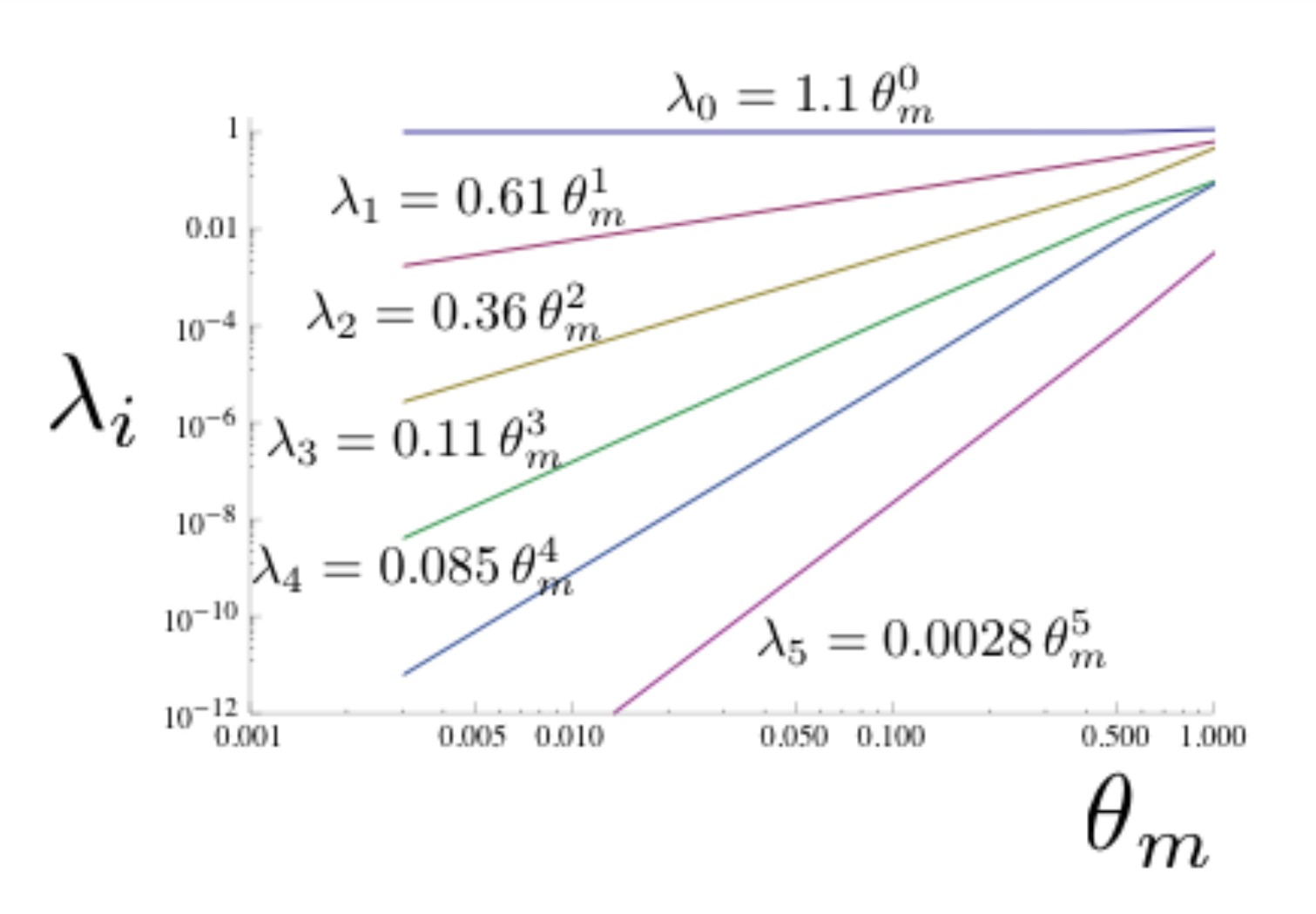}
\caption{\label{fig3} Singular values, $\lambda_i$, as a function of $\theta_m$.}
\end{figure}
\begin{figure}
\noindent\includegraphics[width=15pc]{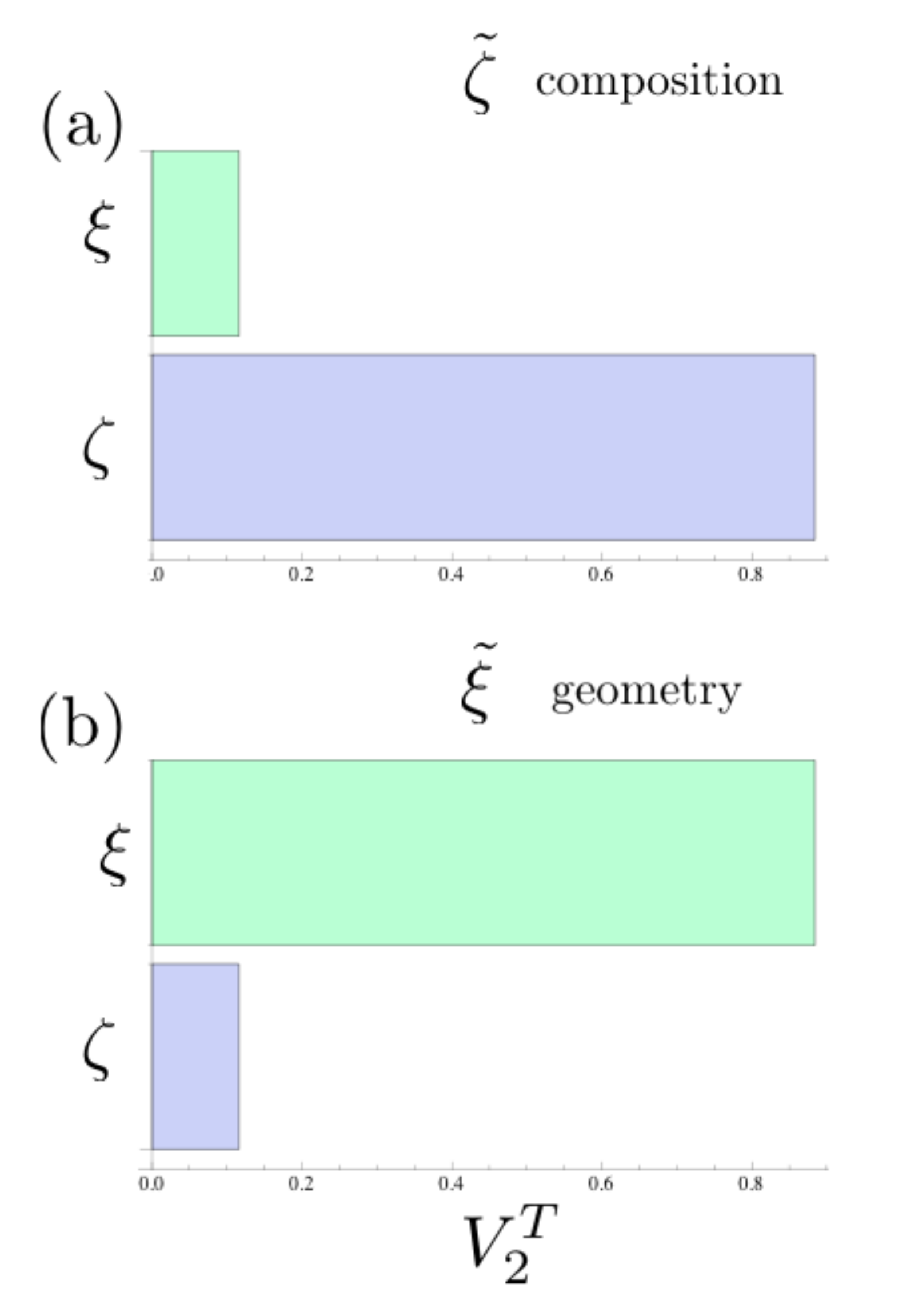}
\caption{\label{fig4} Orthogonal rock physics parameters, $\Delta \tilde{r}$, as given by $V_2^T$.}
\end{figure}
\begin{figure}
\noindent\includegraphics[width=15pc]{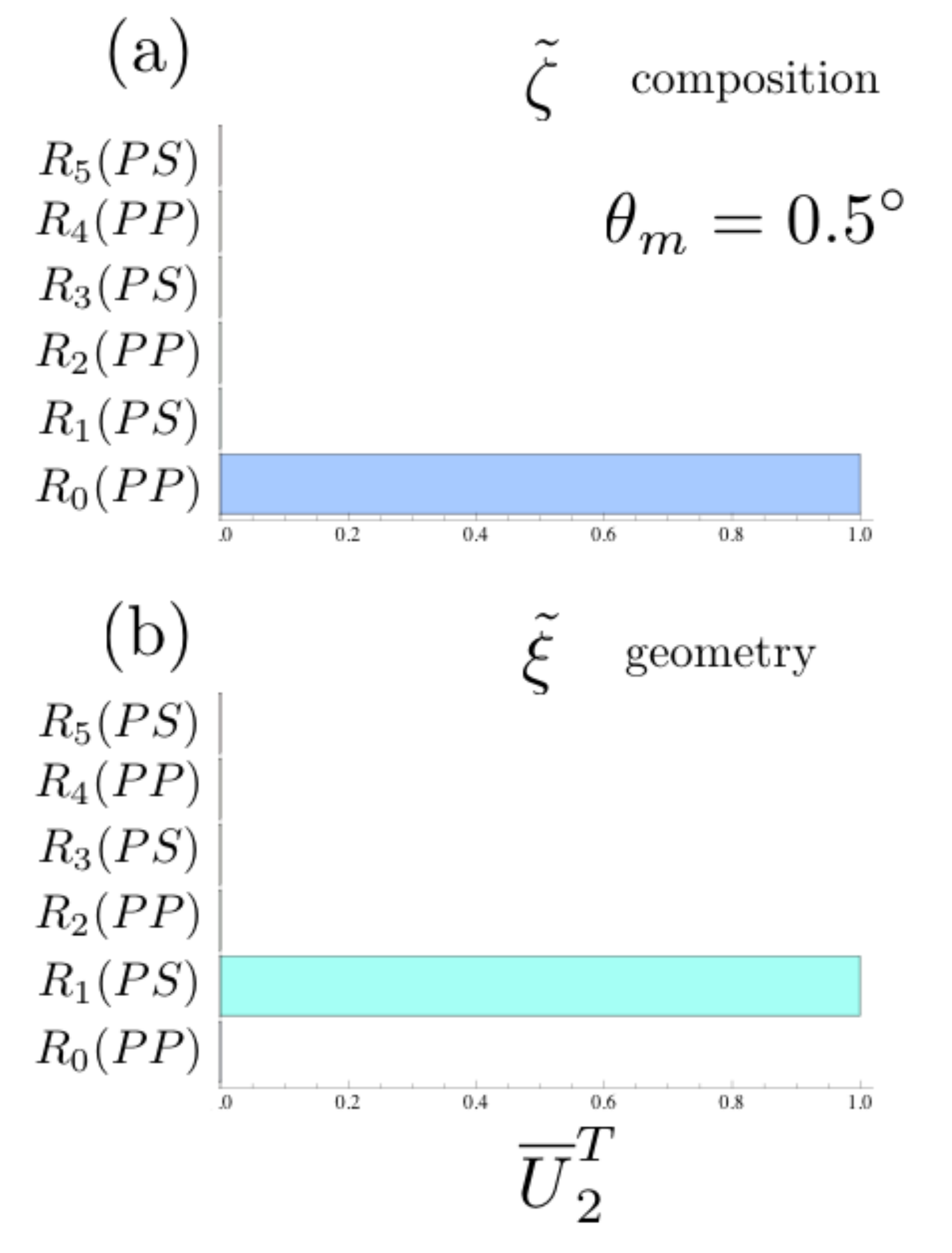}
\caption{\label{fig5} Transformation of the stacks onto the rock physics parameters, $\overline{U}_2^T$, for $\theta_m=0.5^\circ$.}
\end{figure}

\begin{figure}
\noindent\includegraphics[width=15pc]{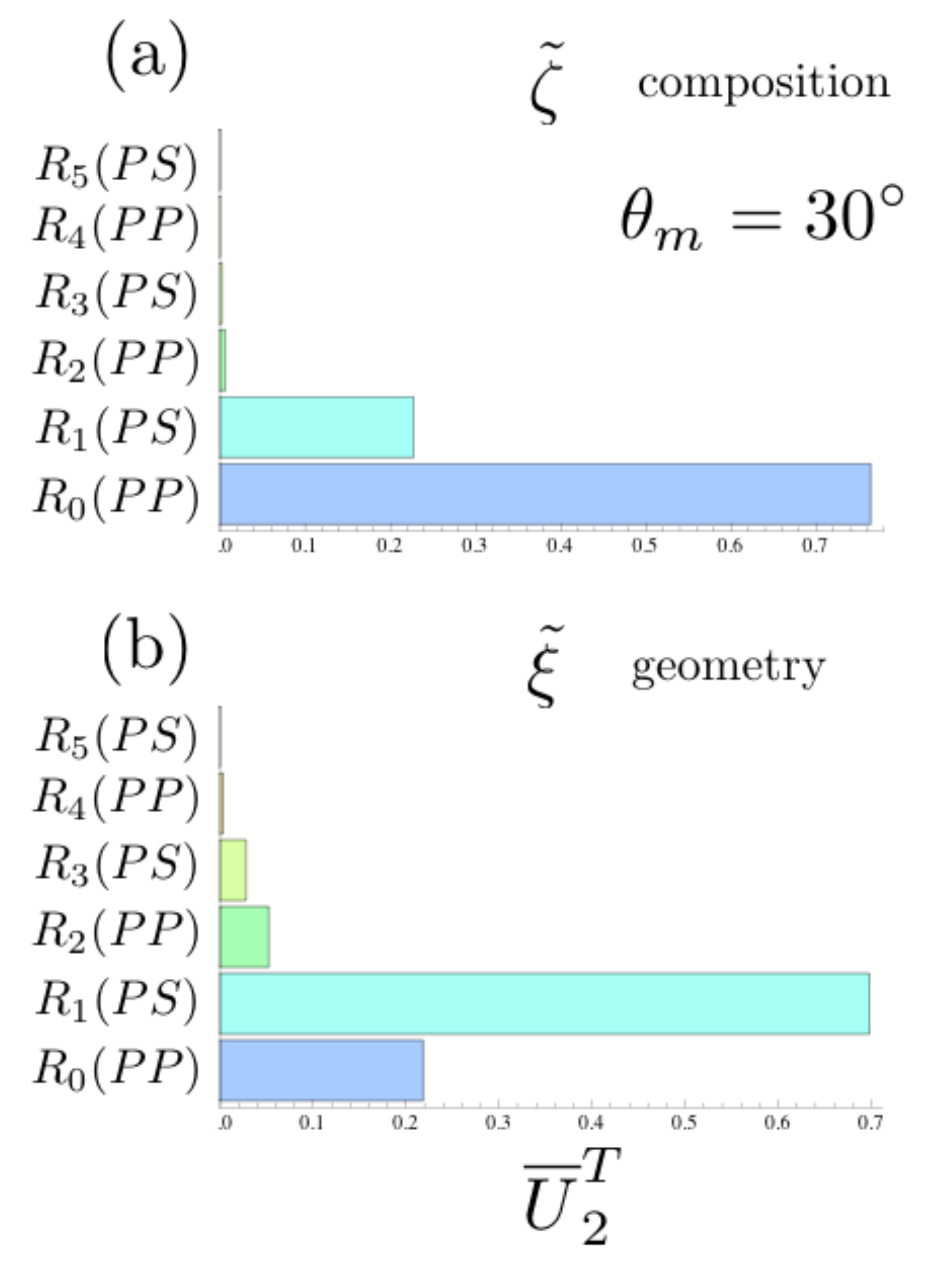}
\caption{\label{fig6} Transformation of the stacks onto the rock physics parameters, $\overline{U}_2^T$, for $\theta_m=30^\circ$.}
\end{figure}

We now increase the maximum angle of incidence to a typical value of $\theta_m=30^\circ$.  The main change is shown in Fig. \ref{fig6} which shows the $\overline{U}_2^T$ transformation.  Although the alignment of the $\tilde{\zeta}$ and the $\tilde{\xi}$ directions stay in the same general directions, they are starting to rotate in the $R_0$-$R_1$ plane (full PP and ``full'' PS) so they are becoming a bit of an admixture of both.  Note that the AVO PP stack, $R_2$, and the 4th order PP stack, $R_4$, still have negligible contribution to both.  The reason for this can be seen in the $\overline{\Sigma}_1$ singular values of the $\tilde{R}$ stacks.  The second singular value, $\lambda_1$ (of the ``full'' PS stack) is 10 dB less than the first singular value $\lambda_0$ (of the full PP stack).  The singular value of the AVO PP gradient stack, $\lambda_2$, is an additional 12 dB less that that of the ``full'' PS stack, so that it is 22 dB less than that of the full PP stack.  It should be noted that the singular value of the 4th order PP stack, $\lambda_4$, is 43 dB less than that of the full PP stack.  Since the expected SNR of most seismic data is 10 dB to 20 dB, one can reasonably expect to reliably estimate the full PP and the ``full'' PS stack.  It is rather tenuous whether the AVO PP gradient stack can be estimated.  There is little probability that the 3-term AVO, as determined by 4th order PP stack, can be estimated reliably.

\begin{figure}
\noindent\includegraphics[width=15pc]{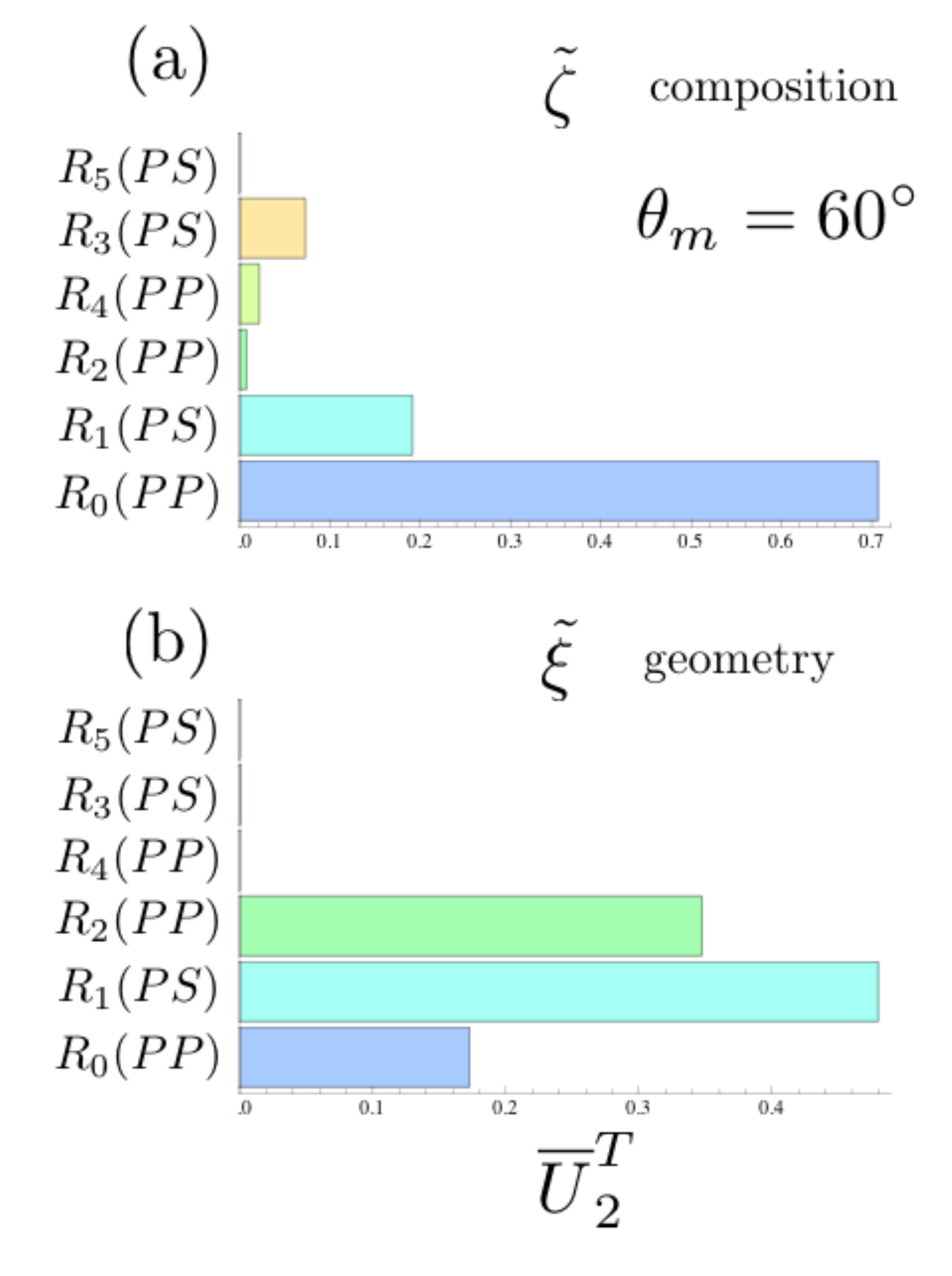}
\caption{\label{fig7} Transformation of the stacks onto the rock physics parameters, $\overline{U}_2^T$, for $\theta_m=60^\circ$.}
\end{figure}

\begin{figure}
\noindent\includegraphics[width=20pc]{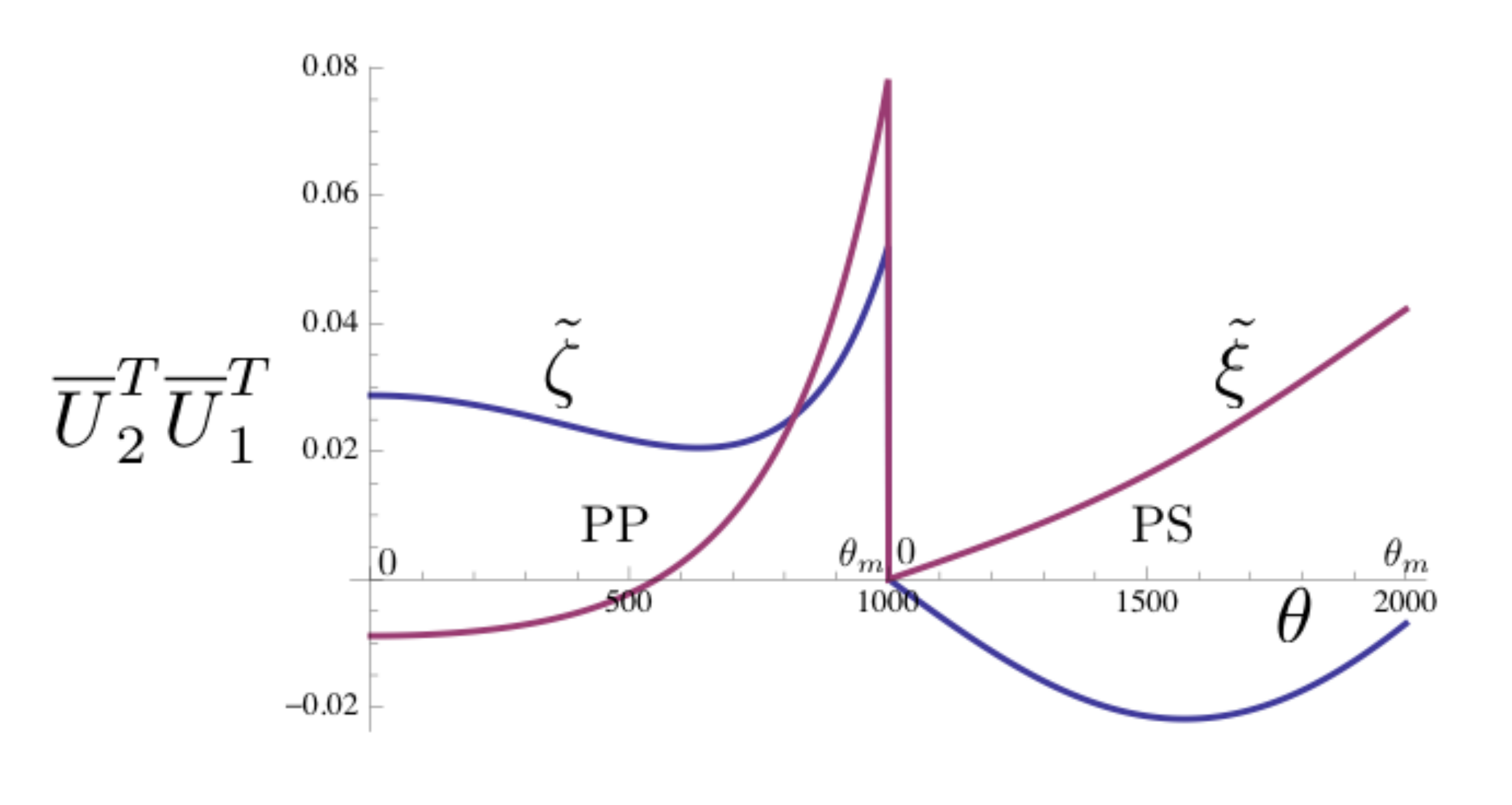}
\caption{\label{fig8} Optimal stack weights, $\overline{U}_2^T \overline{U}_1^T$, as a function of incidence angle, $\theta$, for the determination of rock physics parameters.  First set is for PP data, followed by the weights for PS data.}
\end{figure}

Finally, we increase the maximum angle to $\theta=60^\circ$.  This is representative of very long offset AVO data.  The main change is shown in Fig. \ref{fig7}, which shows the $\overline{U}_2^T$ transformation.  It shows the same modest rotation in the $\tilde{\zeta}$ and $\tilde{\xi}$ directions as the previous case.  The main difference is that the AVO PP gradient stack contributes almost equally with the ``full'' PS stack to the determination of $\tilde{\xi}$.  The reason for this can be seen in the singular values of $\overline{\Sigma}_1$.  The singular value of the ``full'' PS, AVO PP gradient stack, and the 4th order PP stack are 3 dB, 6 dB, and 20 dB less than the full PP stack, respectively.  It is interesting to examine the compound transformation, $\overline{U}_2^T \overline{U}_1^T$, that defines the two optimal stacks for estimation of the two rock physics parameters, $\Delta \tilde{r}$.  They are shown in Fig. \ref{fig8}.  The optimal stack weights for the composition, $\tilde{\zeta}$, are a difference between the full PP and the ``full'' PS stack.  The optimal stack weights for the more important property, the geometry, $\tilde{\xi}$, has roughly equal weights for the ``full'' PS stack, and the far offset PP data.

\begin{figure}
\noindent\includegraphics[width=15pc]{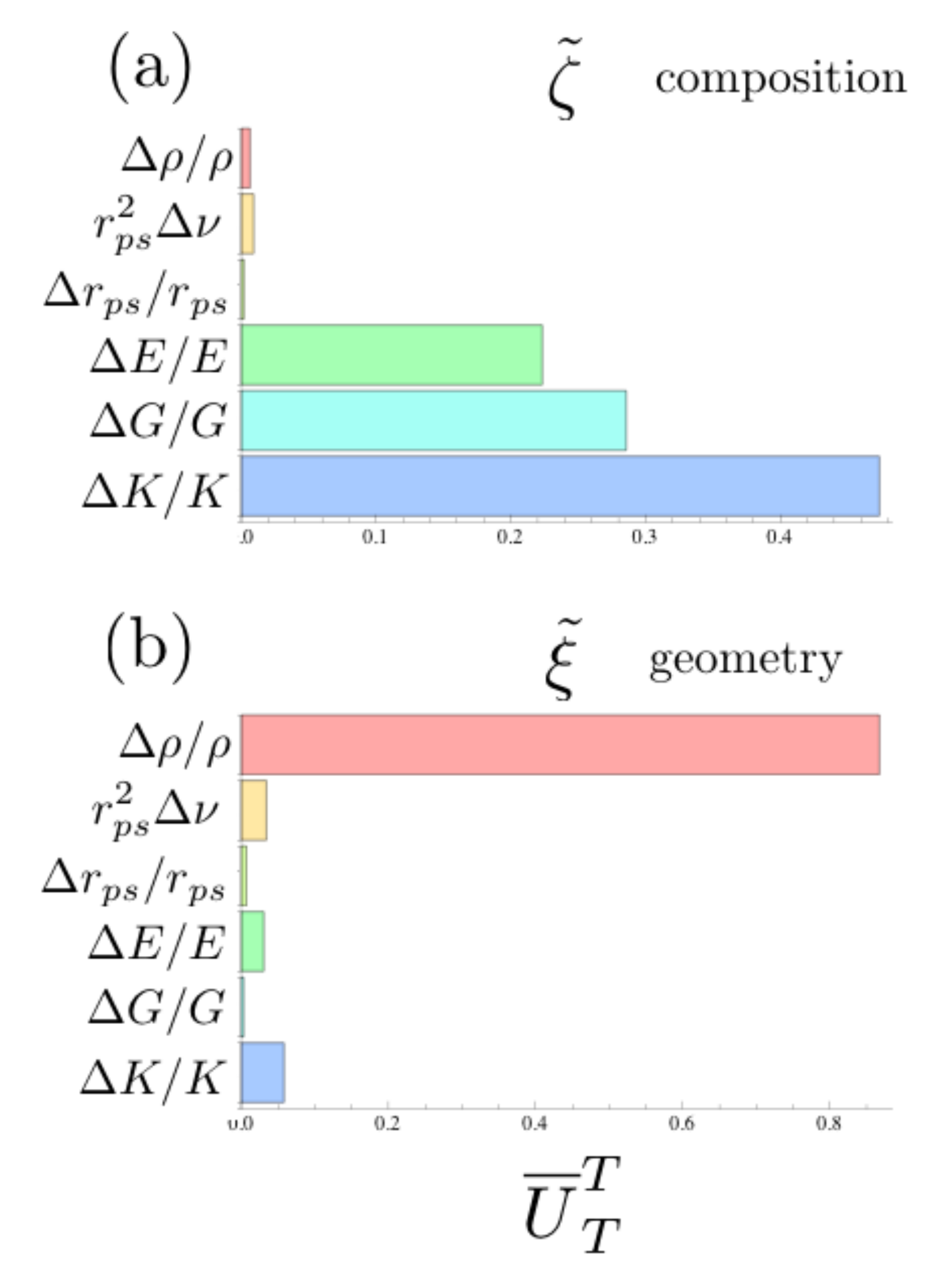}
\caption{\label{fig9} Relationship between traditional rock physics parameters and the fundamental rock physics parameters given by $\overline{U}_T^T$.}
\end{figure}

As we developed earlier, in the theoretical part of the previous section, there is value in examining the relationship between the rock physics and more traditional elastic parameters, $\overline{U}_T^T$.  For the rock physics characteristic of the Marcellus shale, the results are shown in Fig. \ref{fig9}.  All of the moduli, whether the bulk, shear, or Youngs molulus (i.e., $R,G, \text{or} \, E$) have roughly equivalent ability to descern the composition, $\tilde{\zeta}$.  For the geometry, $\tilde{\xi}$, however, it is clearly the density, $\rho$, which is the whole story.  One will need to estimate one of the moduli before the secondary variation (secondary singular value) associated with the density can be understood, though.  We are not advocating inverting for the density.  First of all, it is an absolute property, not a relative property like $\Delta \rho / \rho$.  There are grave technical concerns in inverting for such absolute quantities because of the need to incorporate absolute reference values.  They are never truly known, and incorporation of them in the results will bias the results.  Second, it is an un-necessary complication to invert for a meta parameter, and it complicates the incorporation of prior information.  Instead, one should invert directly for $\xi$ from a limited number of stacks of $\tilde{R}$, where the data covariance is diagonal and largest.  This analysis does confirm, though, some of the folklore that believes it is density that matters in predicting the performance of unconventional reservoir fracturing.

We now turn our attention to how noise and systematic data distortions will modify what the optimal stack weights will be.  In practice, these weights are determined by a principal components analysis of the seismic data.  The renormalization constants, $a_i(s)$, the averaged wavelets, $W_i(t)$, as well as the data covariance matrix, $\Sigma_m$, are also determined by the wavelet derivation process\citep{gunning.glinsky.06} at a well location.  All of these parameters are estimated by a minimization of synthetic seismic mismatch with an additional estimate of the uncertainty in this minimalization.  What we wish to show by this study are reasons for the deviation of the optimal stack weights from the theoretical ones shown earlier in this section.

\begin{figure}
\noindent\includegraphics[width=20pc]{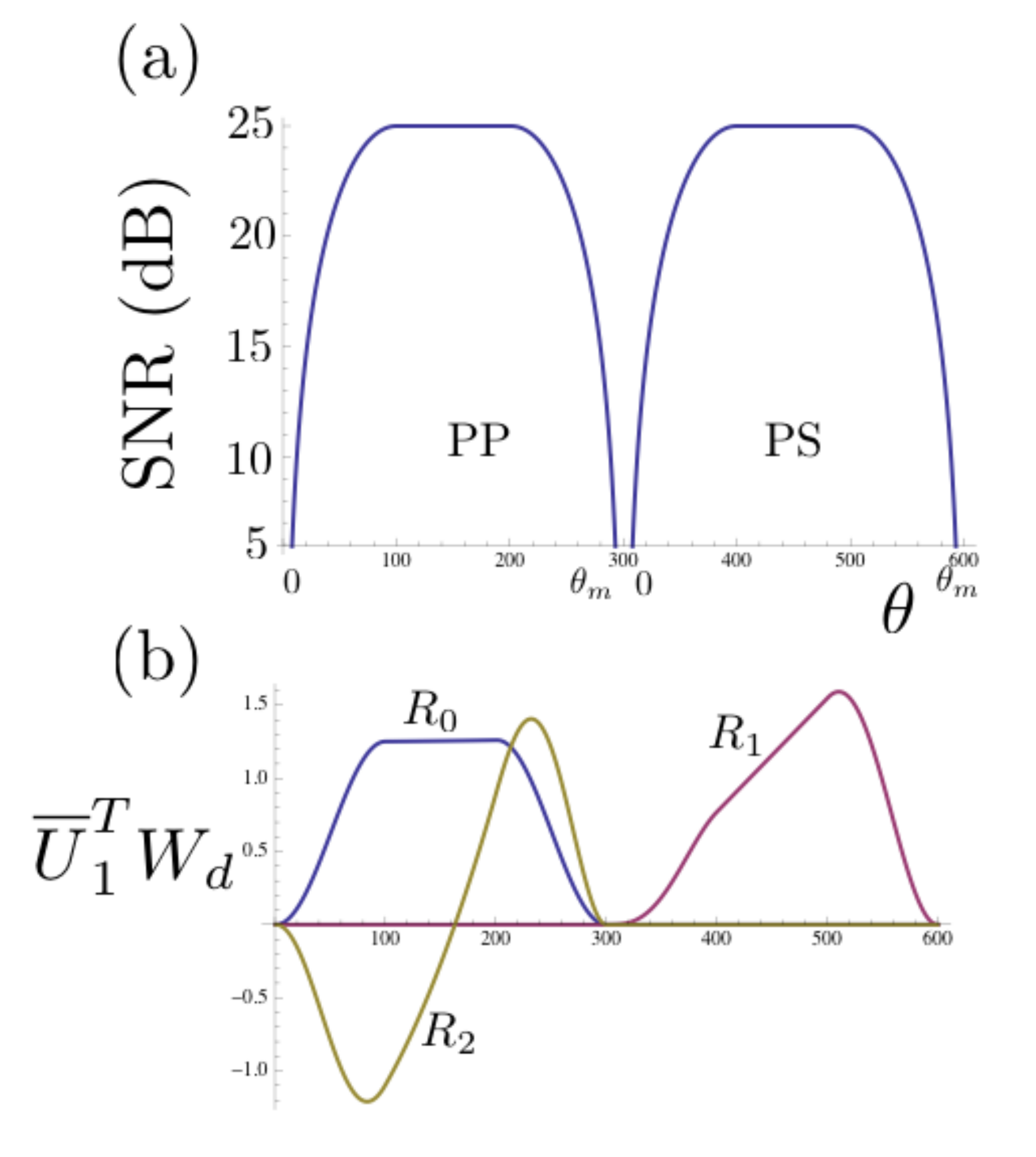}
\caption{\label{fig10} Effect of angle dependent noise on stack weights.  (a) more noise is assumed on the near and far offsets as shown by the SNR, $W_d$, as a function of angle. (b) Stack weights, $\overline{U}_1^T W_d$, as a function of incident angle, $\theta$, for the PP and PS data.}
\end{figure}

We start by showing the effect of having more noise on both the near and far offsets.  The nominal SNR is chosen to be 25 dB.  For simplicity, we have used the three term expression for $M_\theta$ and $M_A$ given in Eq. (\ref{three.term.eqn}).  A diagonal form of $W_d$ is chosen with the diagonal elements shown in Fig. \ref{fig10}a.  The effect on the stack weights, $\overline{U}_1^T W_d$, are shown in Fig. \ref{fig10}b.  They display a common taper that is traditionally applied to weighted stacks at small and large offsets.  This analysis gives a possible physical origin for such a taper.  Such tapers are also found by the principal components analysis discussed in the previous paragraph and the analysis to be shown in Sec. \ref{stack.weights.section}.  The singular values of the stacks, $\overline{\Sigma}_1$, are 23 dB, 12 dB, and -2 dB for the the full PP, ``full'' PS, and the AVO PP gradient stacks, respectively.

\begin{figure}
\noindent\includegraphics[width=20pc]{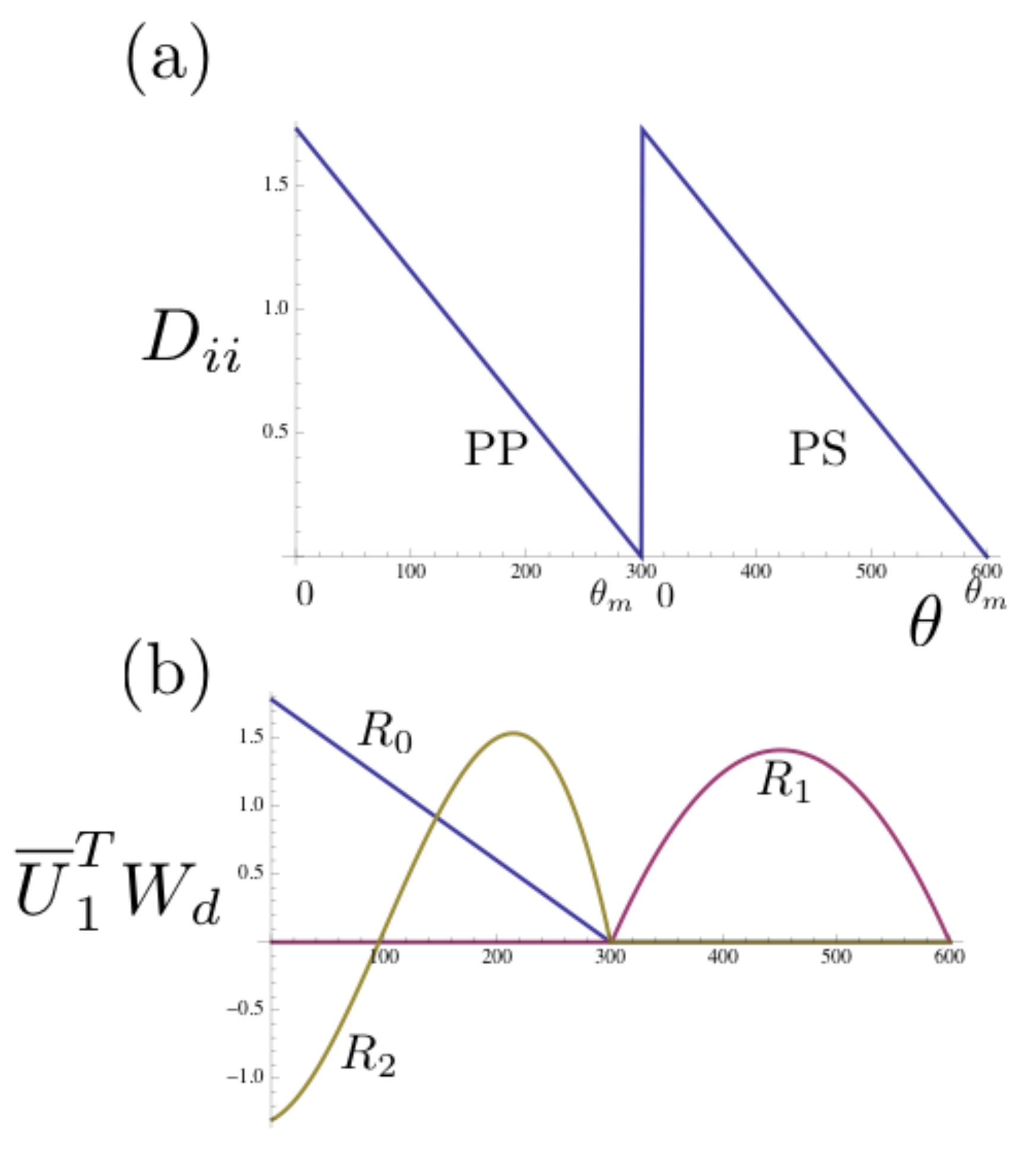}
\caption{\label{fig11} Effect of angle dependent distortion on stack weights, with a constant SNR.  (a) ``hot nears'' such that the near offset traces are artificially enhanced is shown by the offset dependent distortion, $D_{ii}$, as a function of angle. (b) Stack weights, $\overline{U}_1^T W_d$, as a function of incident angle, $\theta$ for the PP and PS data.}
\end{figure}

We now simulate another common data non-ideality -- ``hot nears'', an offset dependent distortion, diagonal $D$, such that the near offset traces are artificially enhanced (see Fig. \ref{fig11}a).  If the SNR after this distortion is a constant 25 dB, the stack weights, $\overline{U}_1^T W_d$, are shown in Fig. \ref{fig11}b.  The effect is counter intuitive.  Since the far offsets have been multiplied by a smaller number, one might expect them to have a larger weight in the stack to compensate.  Instead, they have a smaller weight.  This is because they have a decreased amount of signal with the same noise.  Hence, the effective SNR is less and hence the weight is less.  The SNR for the three stacks are 25 dB, 9 dB, and -3 dB, respectively.

\begin{figure}
\noindent\includegraphics[width=20pc]{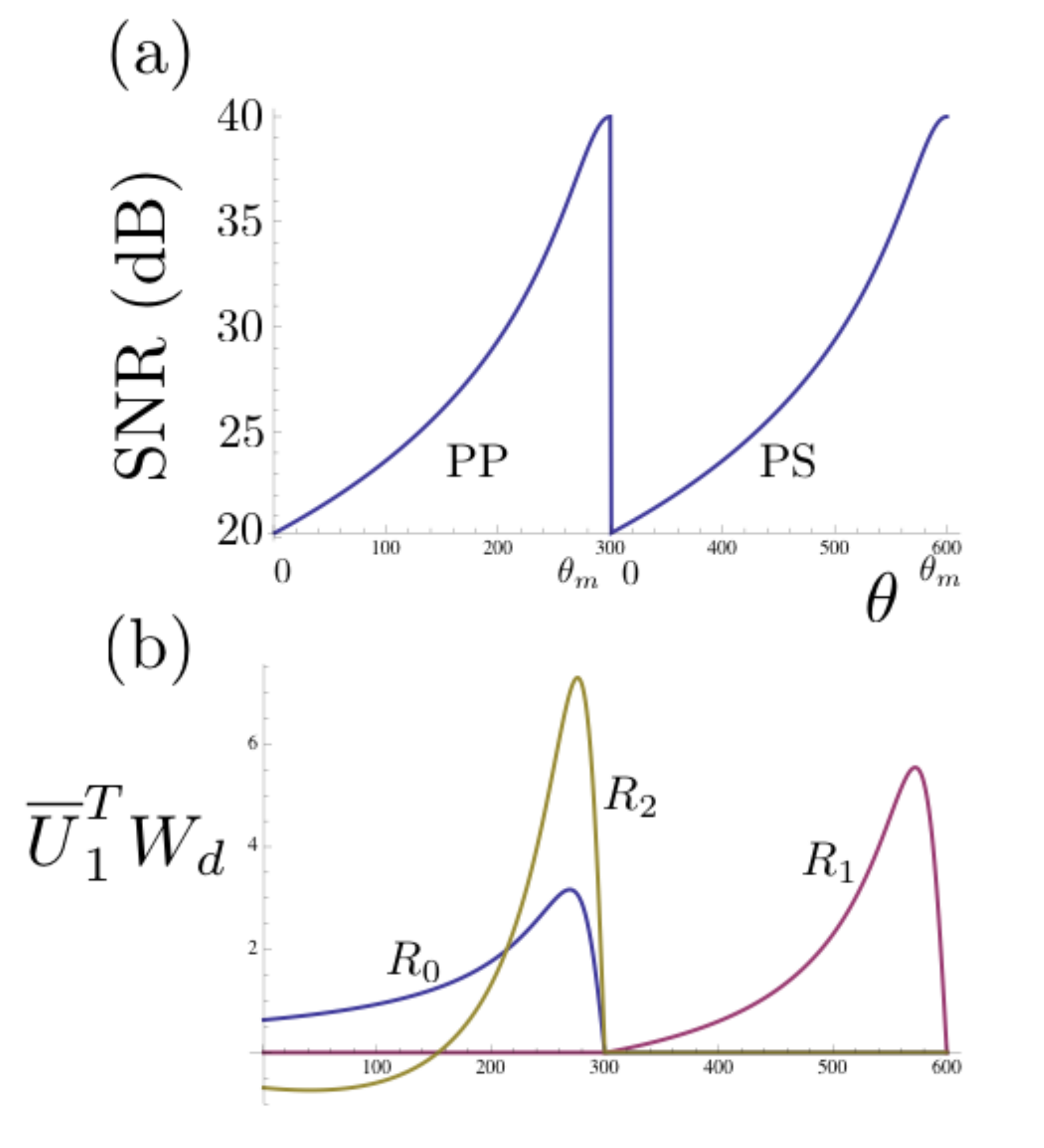}
\caption{\label{fig12} Effect of angle dependent distortion on stack weights, which is applied after the noise.  Angle dependent distortion is the same as that in Fig. \ref{fig11}.  (a) SNR as a function of angle. (b) Stack weights, $\overline{U}_1^T W_d$, as a function of incident angle, $\theta$ for the PP and PS data.}
\end{figure}

Next we assume the same ``hot nears'' of the previous case, but now we assume that the offset dependent scalar is applied after the noise so that the noise level is decreased along with signal.  We limit the SNR to 40 dB.  The same offset dependent weights shown in Fig. \ref{fig11}a are used.  The SNR is modified from a constant 25 dB to that shown in Fig. \ref{fig12}a.  The stack weights $\overline{U}_1^T W_d$ for this case are shown in Fig. \ref{fig12}b.  This result is much more intuitive.  The larger offsets are weighted more to compensate for the smaller multiplicative constant.  This results in the SNR of the second and third stacks to be increased.  The resulting SNRs are 24 dB, 13 dB and 1 dB, respectively.

\subsection{Principal component analysis of stack weights of real data}
\label{stack.weights.section}

The results of Sec. \ref{svd.analysis.section} demonstrated what the theoretical stack weights should be and how angle dependent noise and angle dependent distortions would affect those weights.  Practically, this can be determined from the data.  For some real data characteristic of a typical unconventional shale petroleum reservoir, such an analysis was done on PP data.

A standard principle components analysis was done on the covariance matrix constructed from 12 separate samples of an angle gather.  Each sample has a basis of 40 angles (0 to 40 degrees).  The covariance matrix is $40 \times 40$ and it characterizes at the variance structure of the amplitudes for the 40 angles estimated from our 12 samples.  The eigenvalues and eigenvectors of the covariance matrix are calculated numerically for this square symmetric matrix.  The eigenvalues, or principle components, are proportional to the variance of data associated with the respective eigenvectors.  

\begin{figure}
\noindent\includegraphics[width=20pc]{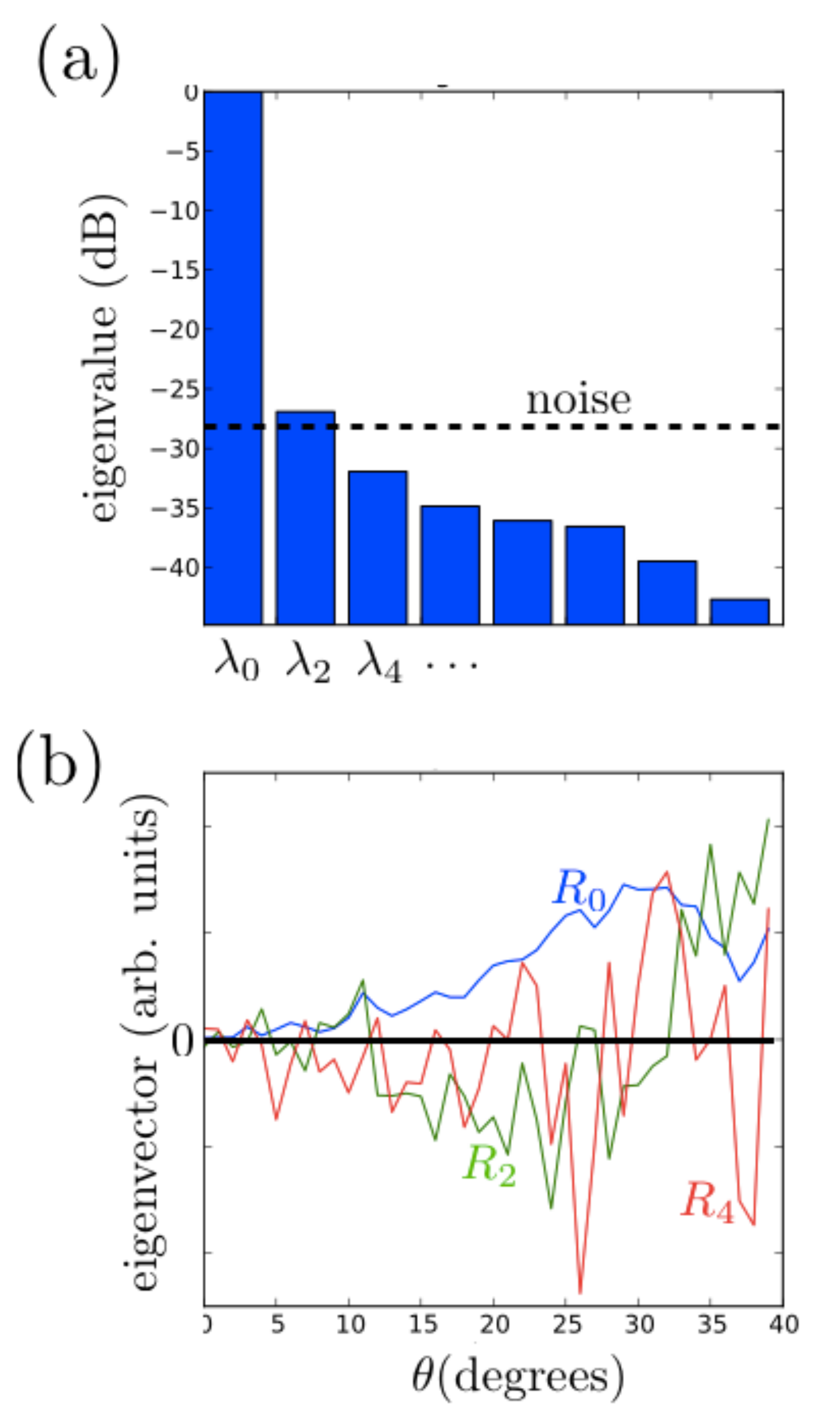}
\caption{\label{fig12a} Results of principal components analysis on real data. (a) eigenvalues displayed in power.  Shown as the dotted line is the noise level as estimated from the form of the eigenvectors. (b) three leading eigenvectors.}
\end{figure}

The results of this analysis are shown in Fig. \ref{fig12a}.  The first eigenvector (labeled as $R_0$ in Fig. \ref{fig12a}b) is smaller than expected for small angles (it should be a constant).  As we have shown in the previous section, this could be because the data has more noise at small angles or because of ``hot nears'' as displayed in Figs. \ref{fig10} and \ref{fig12}, respectively.  We do not know which of these two is the true cause, but we do not need to know.  We just need to form the stacks with these derived weights and proceed with the wavelet derivation process and the rest of the analysis.  The second eigenvector, $R_2$, shows rough characteristics of an AVO PP gradient stack, which is far offsets minus the near offsets.  It does show a large amount of oscillations that have the properties of noise.  This demonstrates that the signal is roughly the same size as the noise.  The third eigenvector, $R_4$, looks like only noise.  This is highlighted in Fig. \ref{fig12a}a, which shows the eigenvalues in reference the the noise level implied by these eigenvectors.

The analysis was continued and a Bayesian wavelet derivation \citep{gunning.glinsky.06} was done using a method that estimated the noise using the well log.  The results showed a good match of the synthetic to the seismic and a reasonable wavelet.  More importantly, when the noise level was compared to the size of the dominate reflections, we determined that the SNR was about 20 dB.  This compares to the 28 dB estimated from the principle components analysis.

\subsection{Detectability including rock physics uncertainty}
\label{detectability.rock.physics.section}

\begin{figure}
\noindent\includegraphics[width=20pc]{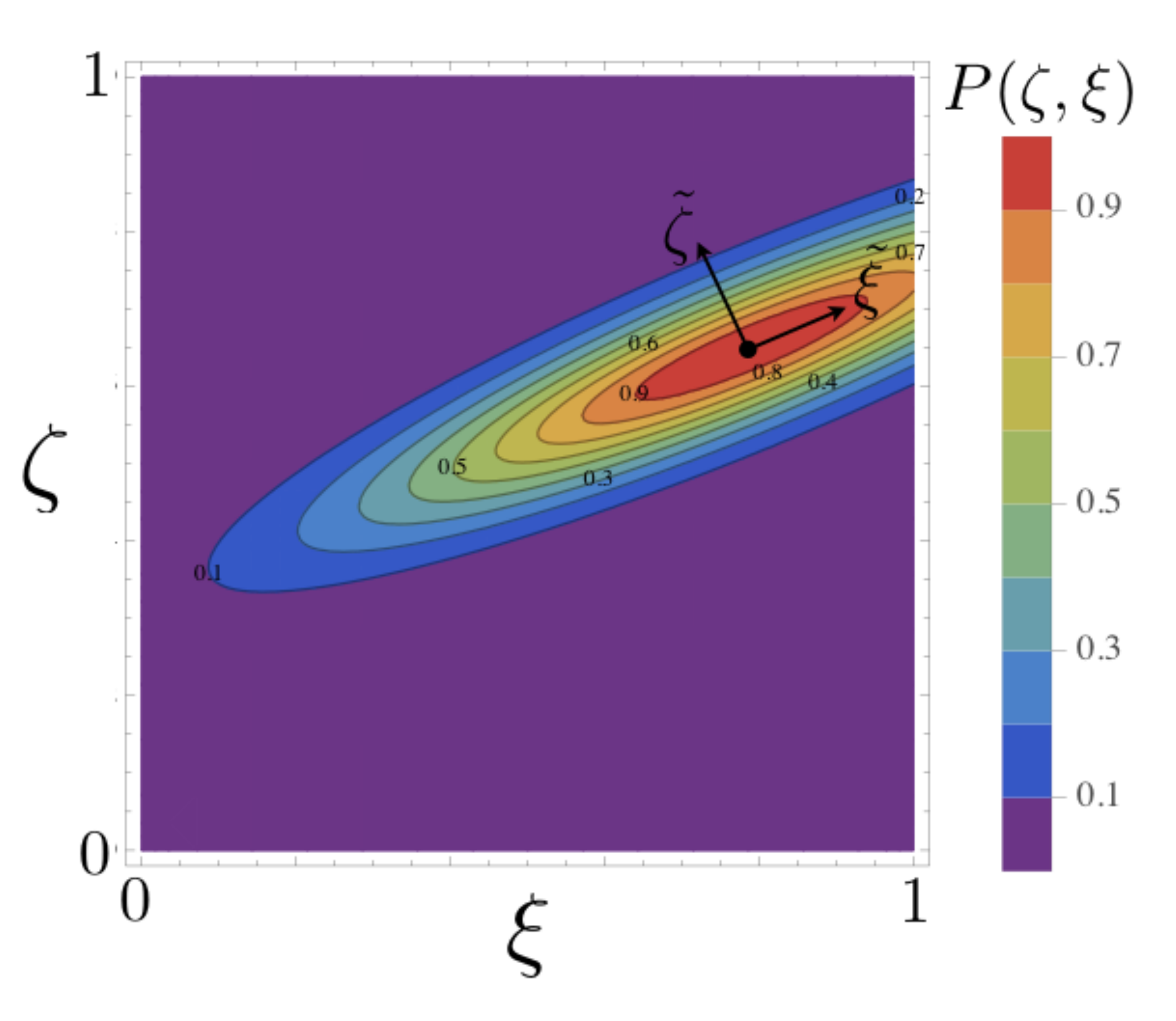}
\caption{\label{fig21} Probability of $r$, $P(\zeta,\xi)$, as a function of $\zeta$ and $\xi$.  The value of $r_1$ that is forward modeled is shown as the black dot.  The principle directions of the distribution are shown as the black arrows.}
\end{figure}

We now turn our attention to the practical detectability of the rock properties.  To do this, we extend the analysis of Sec. \ref{svd.analysis.section} to include the uncertainty in the rock physics, $\varepsilon_r$. We use the expression for the 5 term $A$ vector given in Eq. (\ref{five.term.eqn}), a maximum angle of $\theta_m= 60^\circ$, a data error of 1\% in reflection coefficient (RFC) units, and base values for the rock physics of $r_1=(\zeta_1,\xi_1)=(0.79,0.65)$ characteristic of the Marcellus shale to be discussed in the upcoming Sec. \ref{marcellus.model.section}.  The full probability for $P(r)$ of Eq. (\ref{Pr.eqn}) is shown in Fig. \ref{fig21}.  The untruncated width in the $\tilde{\zeta}$ direction is $0.06$ and is $0.35$ in the $\tilde{\xi}$ direction.  The rotation of the ellipsoid is $23^\circ$.  The dimensions of the ellipsoid is dominated by the rock physics uncertainty for a data error of 1\% RFC.  The data error becomes as important as the rock physics uncertainty in determining the dimensions of the ellipsoids, if it is increased to 3\% RFC.

\begin{figure}
\noindent\includegraphics[width=20pc]{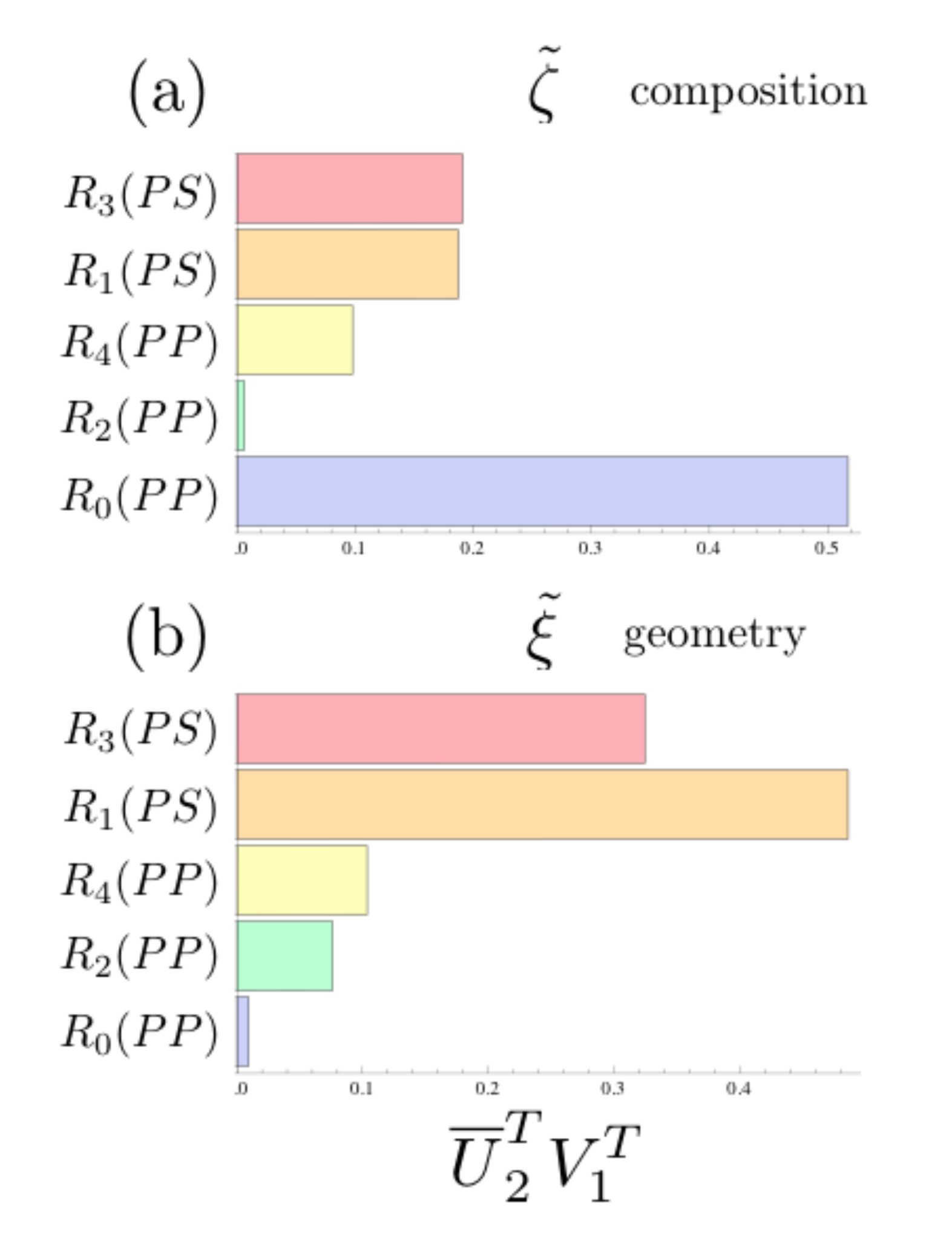}
\caption{\label{fig22} Contribution of each of the stacks to the determination of the principle directions of $P(\zeta,\xi)$.  Display of the elements of the matrix $\overline{U}_2^T V_1^T$.}
\end{figure}

\begin{figure}
\noindent\includegraphics[width=20pc]{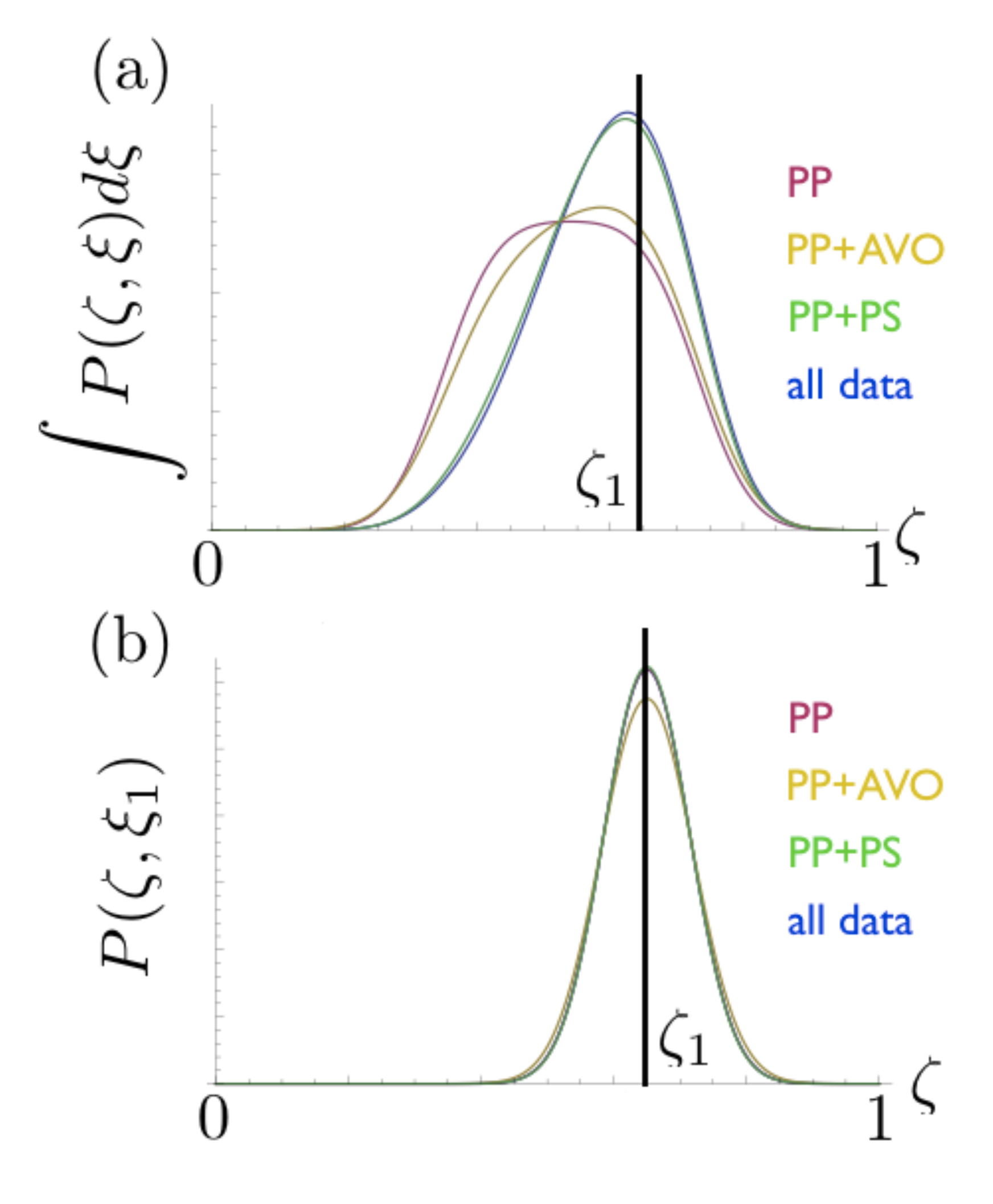}
\caption{\label{fig23} (a) marginal and (b) conditional probabilities of $\zeta$ derived from $P(\zeta,\xi)$.  The true values of $\zeta_1=0.65$ are shown as black lines.  The distribution using the PP data is shown as the magenta line, the PP+AVO data as the yellow line, the PP+PS data as the green line, and all the data as the blue line.}
\end{figure}

\begin{figure}
\noindent\includegraphics[width=20pc]{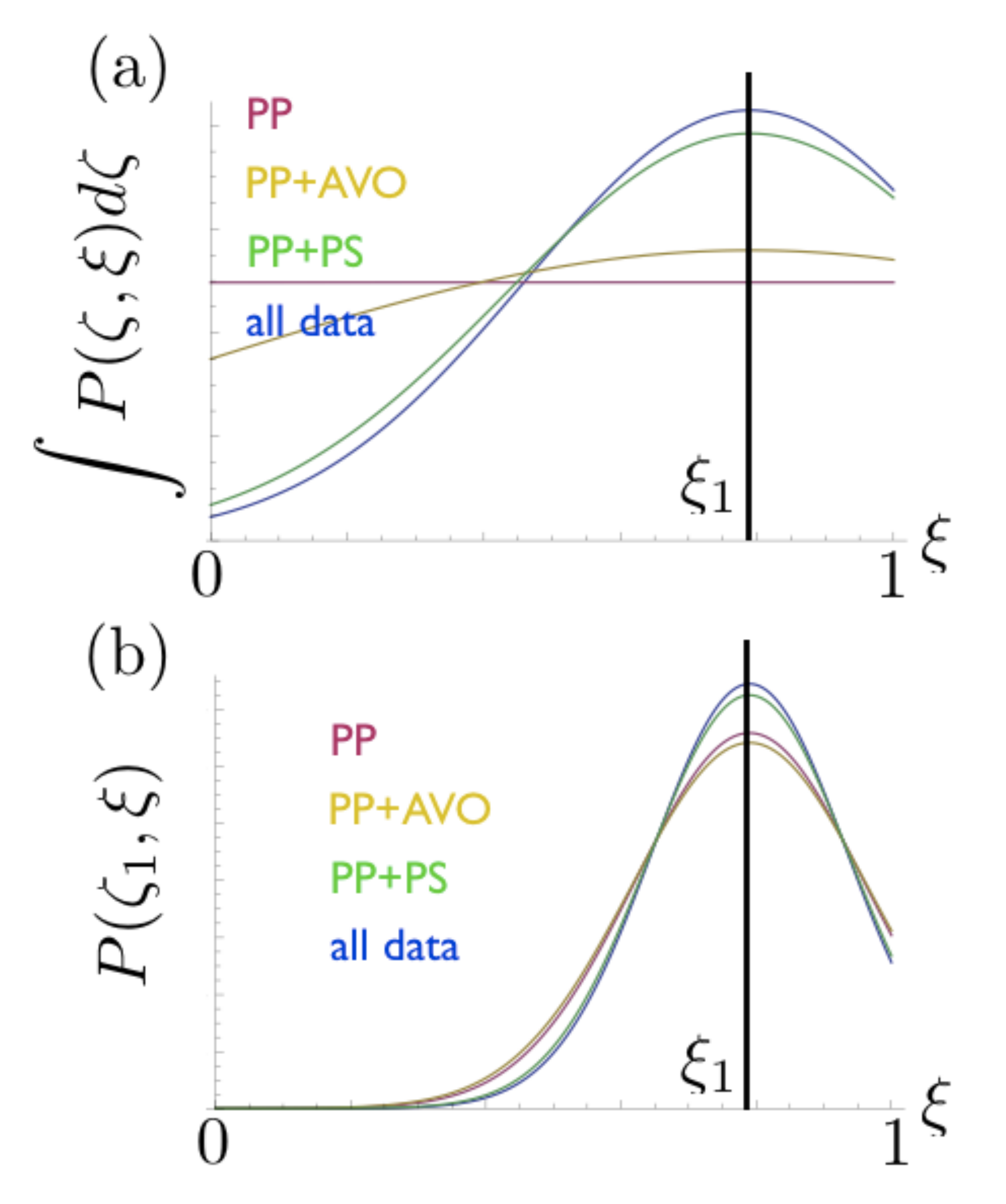}
\caption{\label{fig24} (a) marginal and (b) conditional probabilities of $\xi$ derived from $P(\zeta,\xi)$.  The true values of $\xi_1=0.79$ are shown as black lines.  The distribution using the PP data is shown as the magenta line, the PP+AVO data as the yellow line, the PP+PS data as the green line, and all the data as the blue line.}
\end{figure}

The contribution of each of the terms in the expression for the reflectivity to the determination of $\tilde{\zeta}$ and $\tilde{\xi}$ is shown in Fig. \ref{fig22} by the matrix $\overline{U}_2^T V_1^T$.  The value of $\tilde{\zeta}$ is dominated by $R_0(PP)$ with some contribution from $R_1(PS)$.  The value of the important $\tilde{\xi}$ is dominated by $R_1(PS)$ and $R_3(PS)$.  This is further clarified by examining the marginal and conditional probabilities for $\zeta$ in Fig. \ref{fig23} and for $\xi$ in Fig. \ref{fig24}.  The data that is used (i.e., PP, PP+AVO, PP+PS, or all data) is controlled by manipulation of the data covariance, $\Sigma_m$ (setting the error to a large value for the data to be excluded).  For use of the PP data only, an angle up to $\theta_m=6^\circ$ is used for the PP data.  The marginal probability for $\zeta$ is well determined with a standard deviation of about $0.11$ for all data sets, but a bias of $-0.15$ is removed by including the PS data (the standard deviation is also modestly reduced from $0.13$ to $0.11$).  The conditional probability is well determined for all data sets with a standard deviation of $0.06$.  The marginal probability for $\xi$ is determined with a standard deviation of $0.23$, only with the addition of PS data.  The conditional probability is well determined for all data types with a modest decrease in the standard deviation from $0.14$ to $0.12$ with the addition of PS data.

The optimal stack weights for estimation of $\tilde{\zeta}$ and $\tilde{\xi}$, $\overline{U}_2^T \overline{U}_1^T W_d$, are very similar tho those shown in Fig. \ref{fig8}.  The first set of weights, that estimate $\tilde{\zeta}$, are roughly a full PP plus a ``full'' PS stack.  The second set of weights,  that estimate $\tilde{\xi}$, are a combination of the far offset PS and far offset PP data.

The detectability of the second principle direction, $\tilde{\xi}$, is reduced as the maximum angle is decreased to $45^\circ$, with very little discrimination remaining for maximum offset angles less than $30^\circ$.  The implication is that one can not simultaneously determine $\zeta$ and $\xi$, when the incident angle is under $30^\circ$.  In order to determine $\xi$ for small maximum offset angle, the value of $\zeta$ must be well constrained.  The value of the PS data, in this case, is reduced because the second principle direction is not needed.  However, the value of PS data can be preserved in a multiple layer inversion, at more modest maximum offset angles, as will be demonstrated in Sec. \ref{model.inversion.section}.

\subsection{Marcellus prototype model}
\label{marcellus.model.section}

\begin{figure}
\noindent\includegraphics[width=20pc]{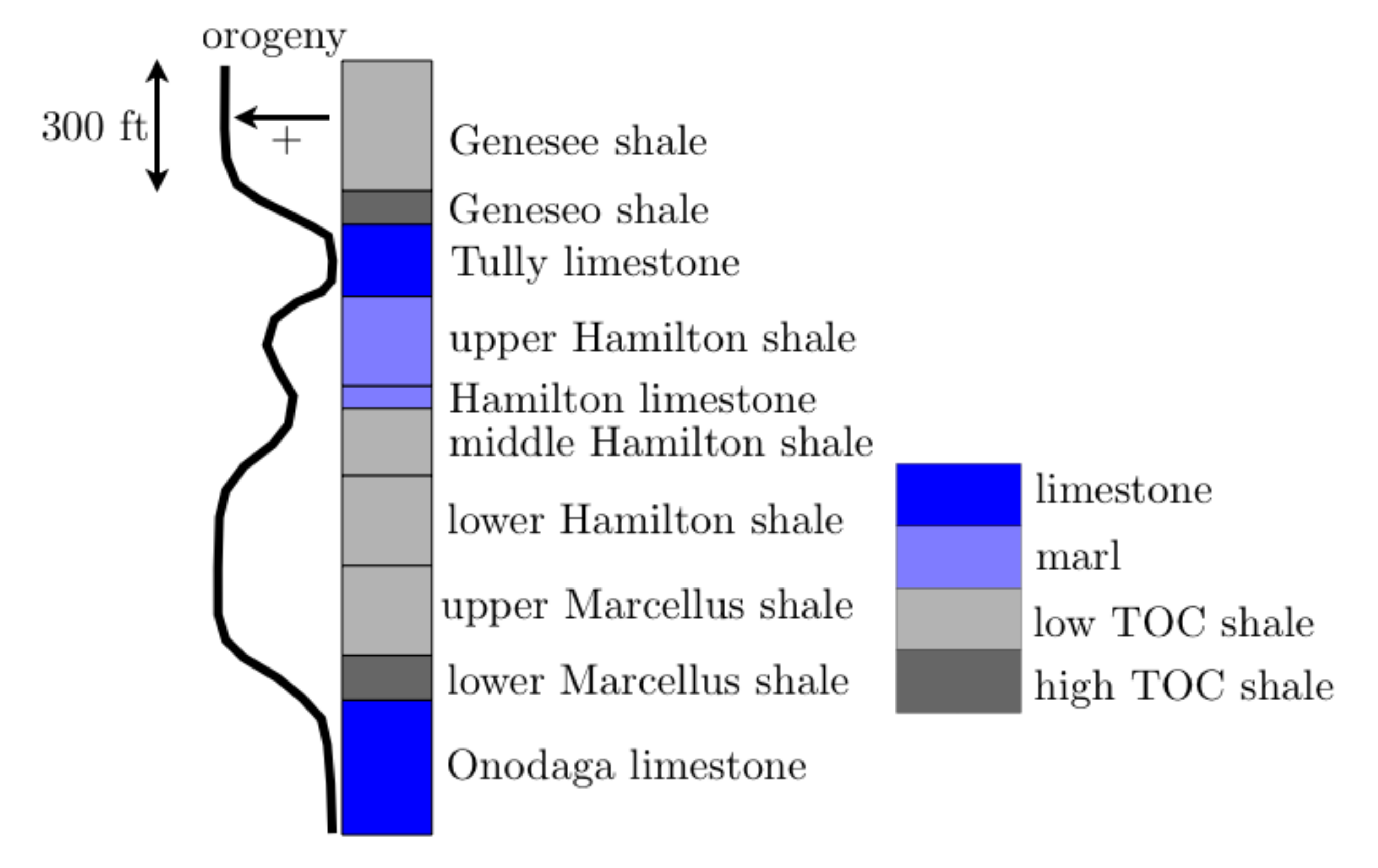}
\caption{\label{fig13} Typical stratigraphic cross section of Marcellus shale play.}
\end{figure}

\begin{figure}
\noindent\includegraphics[width=20pc]{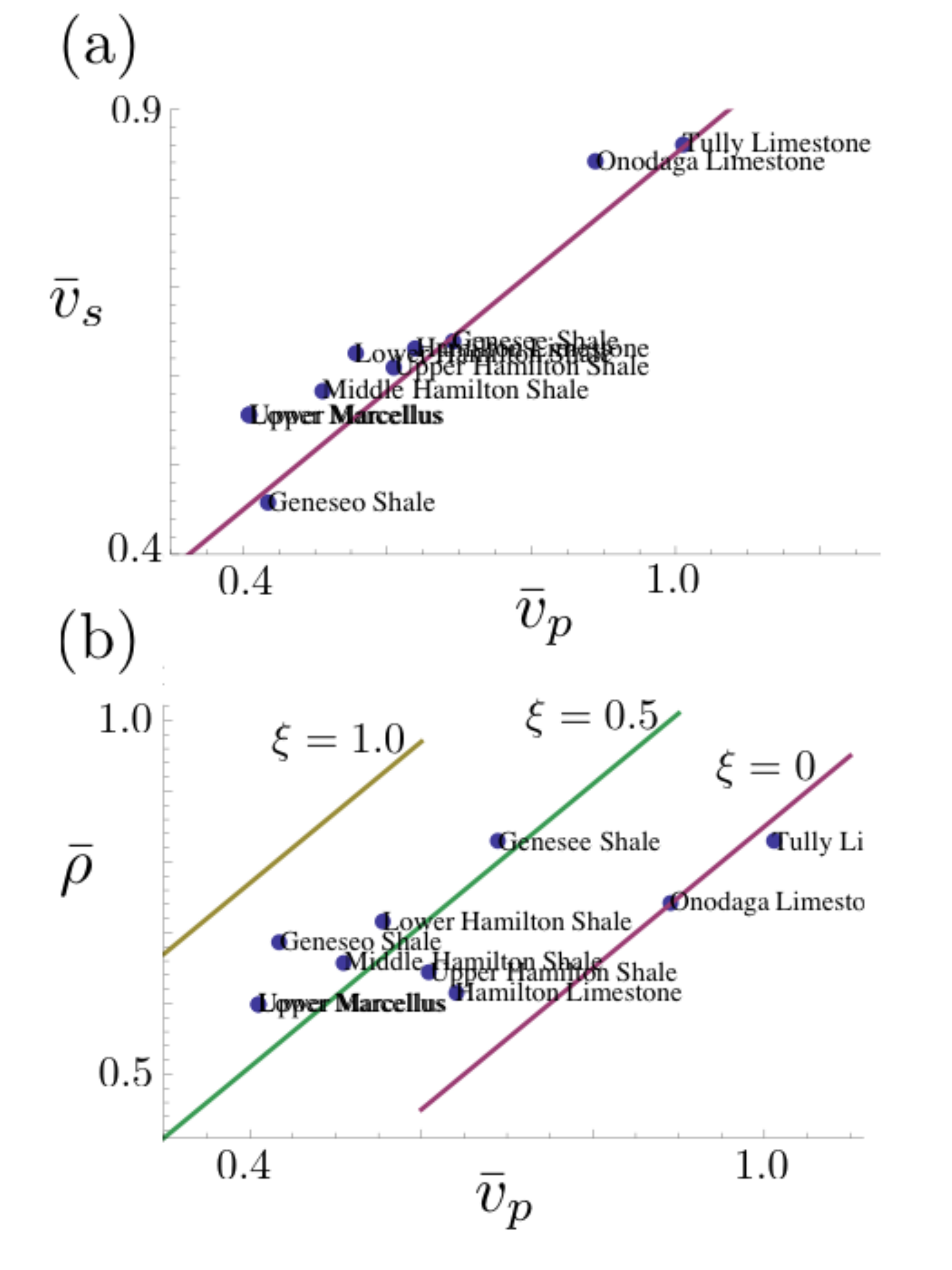}
\caption{\label{fig14} Typical values of $\rho$, $v_p$, and $v_s$ for the Marcellus shale play.  Values are normalized according to the equation $\bar{x} = (x-x_{min})/(x_{max}-x_{min})$, where $\min{v_p}=$ 8000 ft/s, $\max{v_p}=$ 18000 ft/s, $\min{v_s} =$ 3800 ft/s, $\max{v_s}=$ 11000 ft/s, $\min{\rho}=$ 2.1 gm/cc, $\max{\rho}=$ 2.8 gm/cc. (a) $v_s$-$v_p$ values in normalized units.   Purple line is the fit trend, Eq. (\ref{vs.trend.eqn}).  (b) $\rho$-$v_p$ values in normalized units.  Trend lines of constant $\xi$, Eq. (\ref{vp.trend.eqn}), are colored and labeled according to the value of $\xi$.}
\end{figure}

\begin{figure}
\noindent\includegraphics[width=20pc]{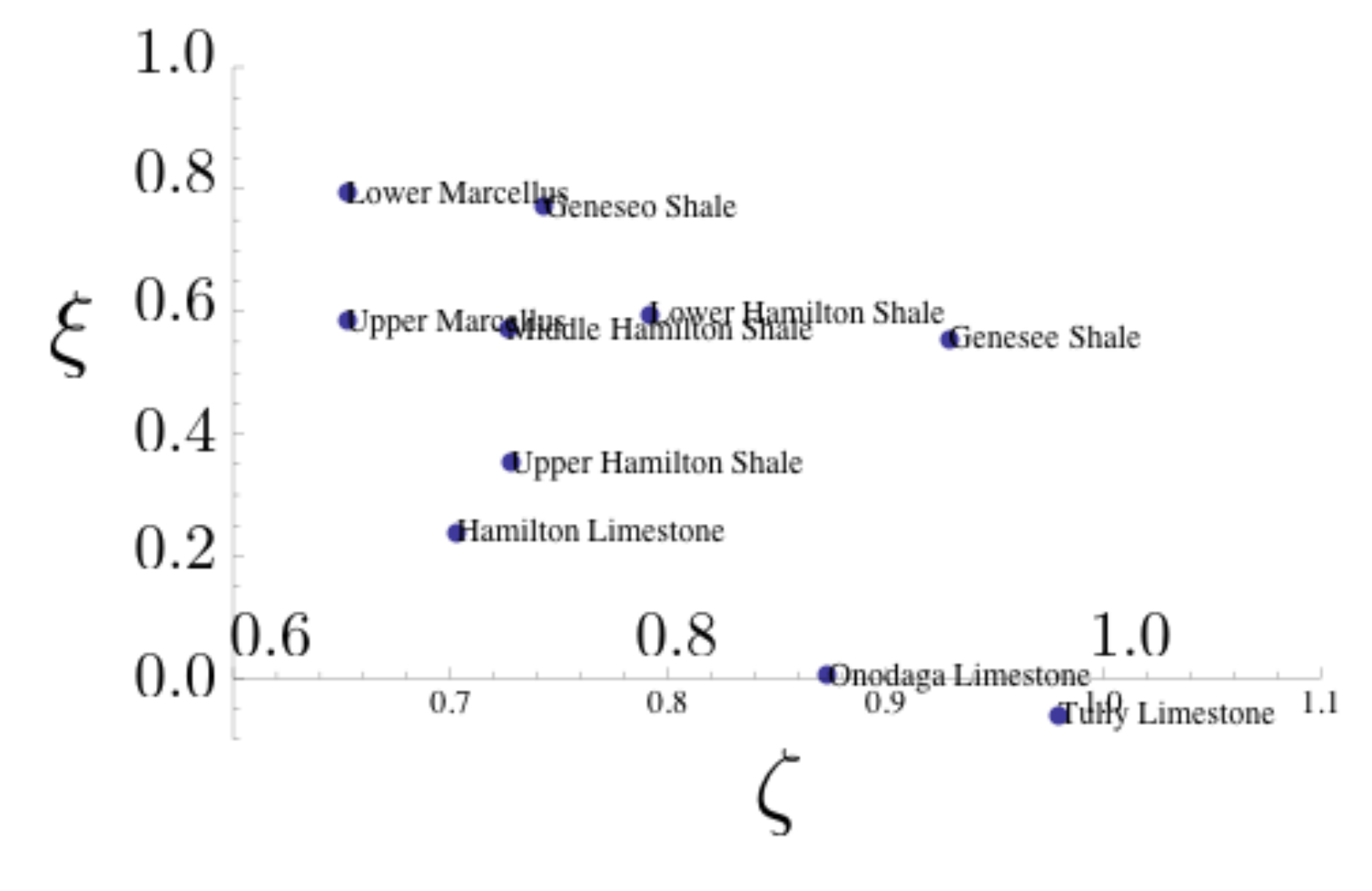}
\caption{\label{fig15} Typical values of $\zeta$ and $\xi$ for the Marcellus shale play.}
\end{figure}

\begin{figure}
\noindent\includegraphics[width=15pc]{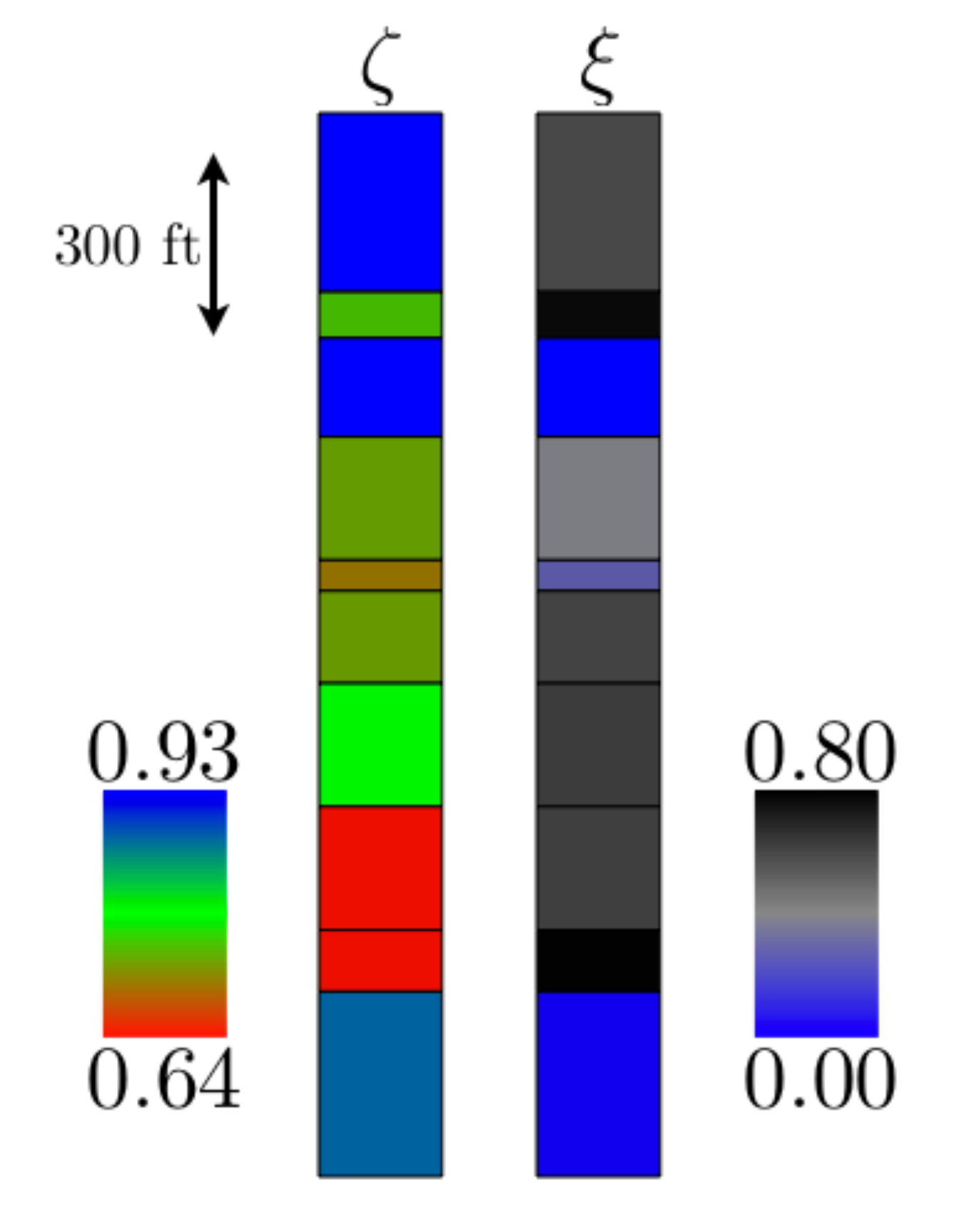}
\caption{\label{fig16} Stratigraphic cross sections with typical values of $\zeta$ (i.e., lithology) and $\xi$ (i.e., compaction, diagenesis, or mineral substitution) for the Marcellus shale play.}
\end{figure}

\begin{figure}
\noindent\includegraphics[width=15pc]{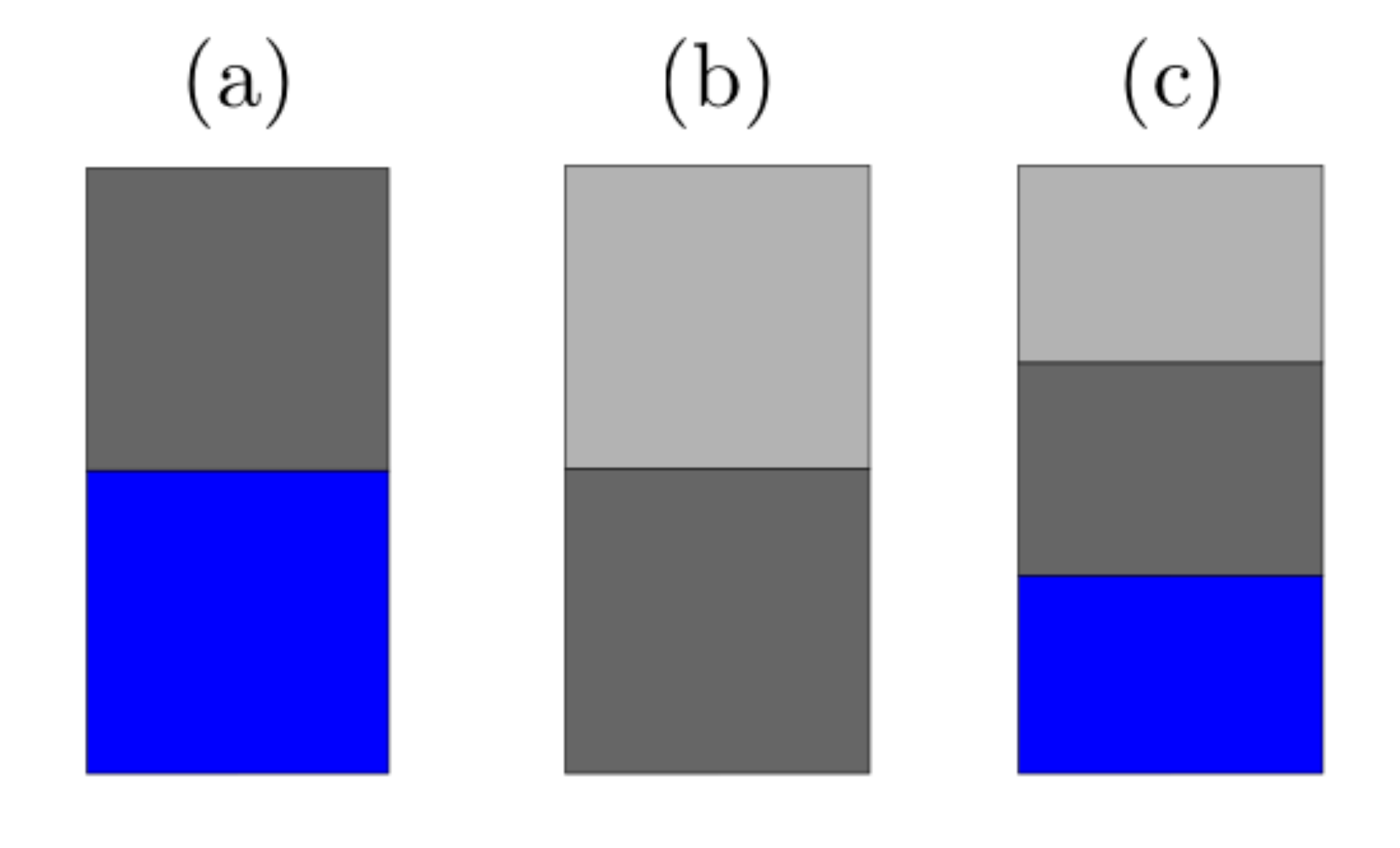}
\caption{\label{fig17} Three simplified models of the Marcellus shale play: (a) high TOC shale on top of a limestone, (b) low TOC shale on top of a high TOC shale, (c) a three layer model of a high TOC shale between a limestone and a low TOC shale.}
\end{figure}

In order to test the practicality of determining the ductile fraction, $f_d = f_{dc} \, \xi$, and other quantities of interest for an unconventional shale petroleum reservoir, a prototype model of the Marcellus play is constructed.  A typical stratigraphic cross section is shown in Fig. \ref{fig13}.  Note that the lower Marcellus shale is the primary interval of interest.  Typical values of $\rho$, $v_p$, and $v_s$ are shown in Fig. \ref{fig14}.  Reference lines of the trends in Eq. (\ref{vp.trend.eqn}) and (\ref{vs.trend.eqn}) are displayed versus these typical values.  The $\rho$ and $v_p$ values are transformed using Eq. (\ref{vp.trend.eqn}) and (\ref{rho.trend.eqn}) to give typical $\zeta$ and $\xi$ for each layer with the results shown in Fig. \ref{fig15}.  Note that the limestones have $\xi \approx 0$, and the marls have $\xi \approx 0.2$.  There are two types of shales.  One type has $\xi \approx 0.5$ and the other type, the high TOC ``frackable'' target shales, has $\xi \approx 0.7$.  The resulting models for $\zeta$ and $\xi$ (where $\xi$ indicates lithology, and $\zeta$ indicates compaction, diagenesis, or mineral substitution) are shown in Fig. \ref{fig16}.  This is consistent with our earlier identification of $\zeta$ with composition, and $\xi$ with geometry.   Three simplified models, two with two layers, and one with three layers, are shown in Fig. \ref{fig17}.  They are contructed to build up to the full model in Fig. \ref{fig16} in a systematic way.  We first will understand what can be learned from the reflection coefficient from the bottom and the top of the target layers in Fig. \ref{fig17}a and Fig. \ref{fig17}b.  The third model (Fig. \ref{fig17}c) adds the additional information of layer times and the accompanying tuning effects.  This is a closer examination of the bottom three layers of the model shown in Fig. \ref{fig13}.

The work of \citet{kohli.etal.13} has shown a strong connection between the ductile fraction and the efficiency of hydraulic fracturing.  For this reason, the main focus will be determining the ductile fraction from the converted wave (i.e., cwave or joint PP and PS) surface imaging of the models of Figs. \ref{fig13} and \ref{fig17}.

It is helpful to understand the geology behind this stratigraphy\citep{sageman.etal.03}.  The Marcellus shale and its accompanying stratigraphy was formed in the Devonian time during the tectonic plate collision that formed the Appalachian mountains.  The deposition was more specifically associated with the foreland basin caused by the isostatic compensation of the thick crust associated with the collision and uplift.  When there was a reduction in the subduction, the sedimentation into the foreland basin was reduced and the basin became shallow enough to be favorable to carbonate formation.  The result was the limestones in the stratigraphic section.  When the orogeny recommenced the basin deepened, but there was a delay in the resumption of the erosion of the mountains and an increase of the sediment load into the foreland basin.  This created a good environment for the formation of shales high in organic content.  As time progressed, the sediment load resumed, increasing the silt in the shale, lowering its organic content and ductile fraction.  This sequence is repeated twice in a significant way in this section, and once in a more minor cycle (see the orogeny curve in Fig. \ref{fig13}).

\subsection{Model based inversion}
\label{model.inversion.section}

It seems difficult to determine the ductile fraction using data with modest maximum angles of incidence, $\theta _m$, for simple two layer models as discussed in Sec. \ref{detectability.rock.physics.section}, and angle dependent wavelet effects with the associated angle dependent tuning.  The complex model described in the previous Sec. \ref{marcellus.model.section}, gives an opportunity to still be successful.  There are advantages introduced by the extra data associated with the multiple reflectors (times and reflection strengths), multiple stacks, differential tuning of the different stack bandwidths, and rich prior model assumptions both on the rock physics and structure.  In order to take advantage of this, a Bayesian model-based inversion\citep{gunning.glinsky.04,chen.glinsky.13} is done.

To test these ideas, a realistic synthetic seismic forward model of the two layer model of Fig. \ref{fig17}a, and the ten layer model of Fig. \ref{fig16} is made.  The ductile fraction rock physics model of Eq. (\ref{vp.trend.eqn}), Eq. (\ref{vs.trend.eqn}) and Eq. (\ref{rho.trend.eqn}) is used.  Uncertainties are assumed to be $25$ m on the thicknesses, $3$ ms on PP times for the bright reflectors, and $8$ ms on PP times for the dim reflectors.  No uncertainty in $\zeta$ is used although the results are relatively unchanged for uncertainty in $\zeta$ less than $0.15$.  The uncertainty in the $\xi$ value is set to $0.20$ except for the limestone layers, which are assumed to have no uncertainty in $\xi$.  The noise on the data stacks is assumed to be 1\% RFC with a maximum offset angle of $\theta_m = 45^\circ$.
\begin{figure}
\noindent\includegraphics[width=20pc]{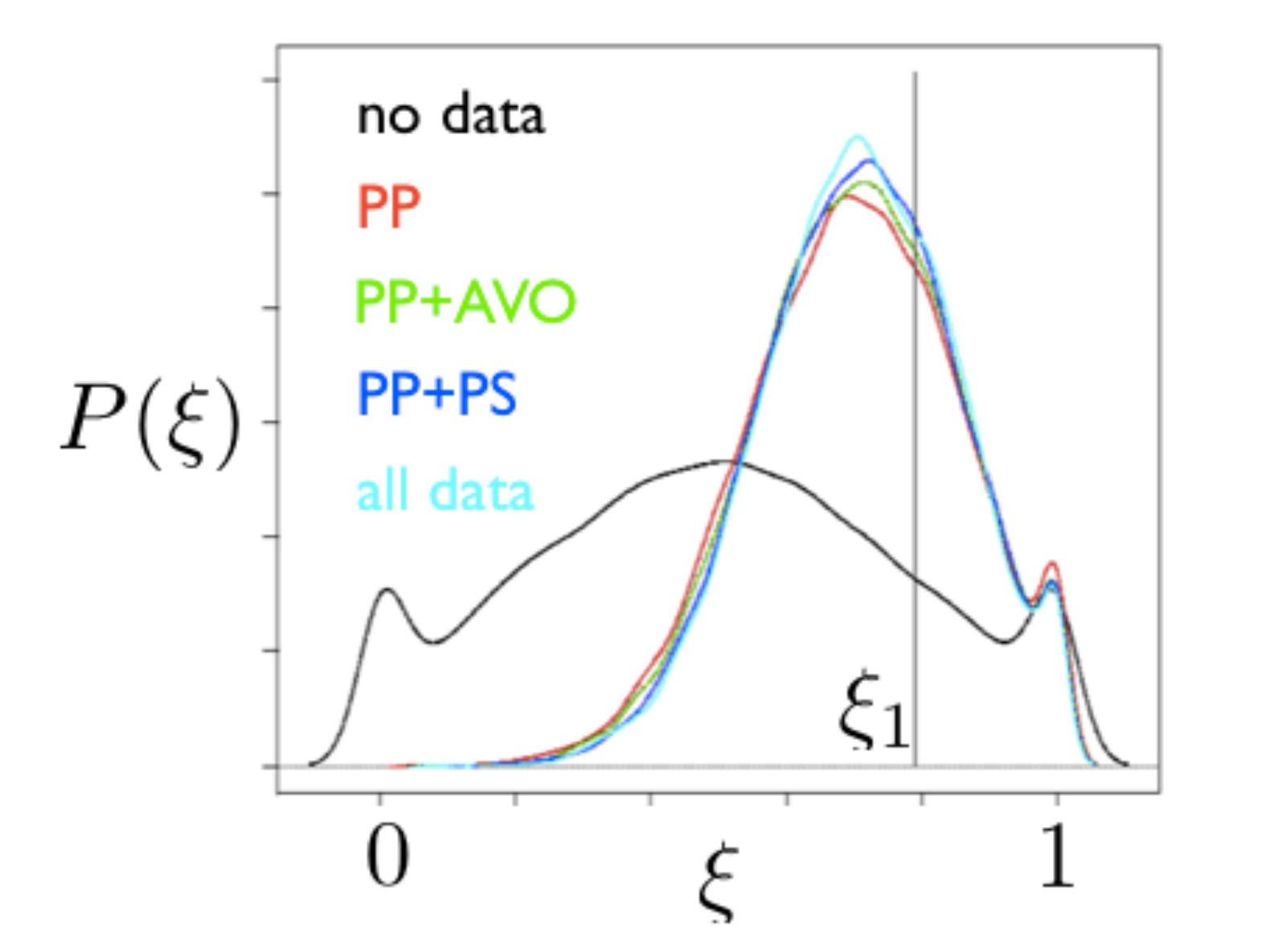}
\caption{\label{fig25} Probability distribution of $\xi$ in the overlying shale layer of the two layer model shown in Fig. \ref{fig17}a.  The value of $\xi_1$ forward modeled is shown as the black vertical line.  The value of $\xi_1$ forward modeled is shown as the black vertical line.  The distribution before the use of any seismic data is shown as the black line, after using the PP data is shown as the red line, after using the PP+AVO data as the green line, after using the PP+PS data as the blue line, and after using all the data as the cyan line.}
\end{figure}
\begin{figure}
\noindent\includegraphics[width=20pc]{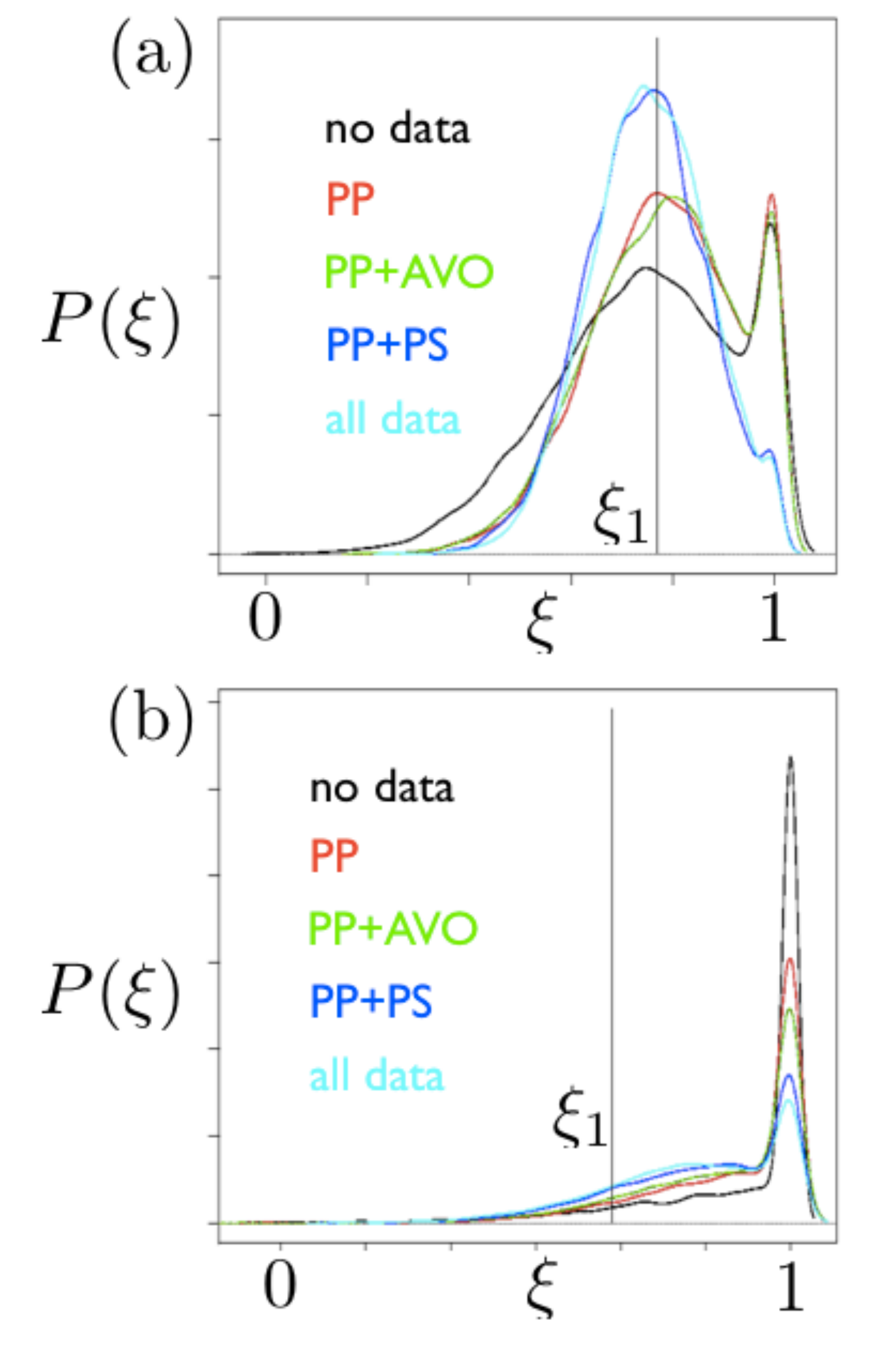}
\caption{\label{fig26}  Probability distribution of $\xi$ in the (a) Geneseo shale and (b) lower Marcellus shale of the ten layer model shown in Fig. \ref{fig16}.  The value of $\xi_1$ forward modeled is shown as the black vertical line.  The distribution before the use of any seismic data is shown as the black line, after using the PP data is shown as the red line, after using the PP+AVO data as the green line, after using the PP+PS data as the blue line, and after using all the data as the cyan line.}
\end{figure}

The results for the two layer model are shown in Fig. \ref{fig25} which displays the probability distribution of $\xi$, $P(\xi)$, for the overlying shale layer.  Note the significant update to the distribution for each seismic data type used and the very modest improvement with the addition of PS data. Both are consistent with the theoretical result of Fig. \ref{fig24}b.  Things become more interesting with the additional complexity and information of the ten layer model.  Figure \ref{fig26} shows the estimated probability of ductile fraction, $P(\xi)$, in the Geneseo shale and the lower Marcellus shale. They represent two situations, one having priors consistent with data (Fig. \ref{fig26}a) and the other having biased priors (Fig. \ref{fig26}b). For the Geneseo layer, without using PS data, the estimated distributions (red and green curves) are bimodal. However, the inclusion of PS data (blue and cyan curves) significantly improves the estimate of ductile fraction, and the unique modes of the posterior distributions correspond to the true value. For the lower Marcellus layer, where the prior is biased to a high value (i.e., $\xi=1.0$), the use of seismic data generally shifts the distributions towards to the true value, and the inclusion of PS data shifts them much more.

\section{Conclusions}
\label{conclusions}

The purpose of this paper has been to establish the underlying fundamentals of quantitative interpretation for unconventional shale reservoirs.  This starts with the understanding that it is the ductile fraction that controls the geomechancial balance between the rocky road joint friction of the fractures and the viscous joint friction.  While this geomechanics is not the subject of this paper, others\citep{zobak.etal.12,kohli.etal.13} have found sensitive dependance of the dynamic friction on the ductile fraction, and a resulting dramatic change in the fracturing efficiency.

Inspired by this geomechanical observation, we developed and verified, at the mesoscopic level, a rock physics model where the three isotropic elastic properties are only a function of two parameters, the scaled ductile fraction, $\xi = f_d / f_{dc}$, and a composition variable, $\zeta = 1 - \exp{(-E / E_0)}$, which captures compaction, diagenesis, and mineral substitution effects.  The first variable captures changes in the geometric microstructure, that is how efficient the rock matrix is in supporting stress -- modulus per mass or coordination number.  The second variable captures the compositional properties of the matrix.  It is a remarkble gift of nature that there are only two parameters and that one of them is directly related to the ductile fraction -- the critical parameter for the geomechanics.

The next important question that we answered is how this geometry parameter, $\xi$, manifests itself in surface reflection seismic.  The equations relating the rock physics to the reflectivity were all linearized and an SVD analysis was done to answer this question.  The leading order singular value was primarily related to the full PP stack and the composition, $\zeta$.  The next order singular value was primarily related to the ``full'' PS stack and the geometry variable, $\xi = f_d / f_{dc}$.  For reasonable angles of reflection, the higher order stacks, which include the AVO PP gradient stack, all have small SNRs which would make them hard, if not impossible, to detect.  If the angles of incidence could be extended to $60^\circ$ or more, the AVO PP gradient stack could be substituted for the ``full'' PS stack because its singular value becomes roughly equal.  We wish to emphasize that this is not a three term AVO analysis for a determination of $\rho$, $v_p$ and $v_s$.  It is only a two term analysis for the two rock physics parameters.  Because the rock physics only has two parameters, using three stacks creates an over determined system.  While using the three stacks would improve the estimates of those two parameters, the third stack is not necessary.

A further analysis was done to relate the two fundamental rock physics parameters to traditional elastic parameters.  It was found that the composition, $\zeta$, is related to the moduli (bulk, shear or Youngs), and the geometry, $\xi=f_d / f_{dc}$, is related to the density.  This is consistent with common wisdom of the density being needed to predict frackability.

There are two other practical findings of this analysis.  The first is that effective wavelets (for each stack) can be used for the first three stacks (i.e., full PP, ``full'' PS, and AVO PP gradient).  This is because corrections to these wavelets would be of higher order (fourth order in $\theta_m$, compared to the second order accuracy of the reflectivity calculation).  Second, the effect of scale can be captured in renormalization constants that are absorbed into the wavelet normalizations and the effective angle of incidence.  These practical findings enable a wavelet derivation process which finds a separate wavelet for each stack and a constant which relates the effective angle of incidence to the true angle of incidence.

The effect of noise that is a function of the angle of incidence, and distortions to the data that are functions of angle of incidence, were shown to be corrected by modification of the stack weights.  These weights are conveniently derived from real data by a principle component analysis on the real data.  The result is the taper at small and large offsets, and an offset dependent scalar being applied to the data.  This analysis gives theoretical justification to common practices that have been done for more practical reasons.

The final portion of this work focused on the practical application of the theory to both synthetic and sometimes to real data.  First, the result of determining the stack weights via a principal components analysis on real data was shown.  The results support the analytic work and the conclusions of that work.  The preliminary SVD analysis was then extended to include rock physics uncertainty and to understand the detectability of ductile fraction.   The results support a detectability of ductile fraction using PS data or large offset PP data.

Finally, a set of synthetic models were constructed that are a realistic reproduction of the stratigraphy and rock physics of the Marcellus shale play.  These models included uncertainty in the rock physics, angle dependent wavelet effects, seismic noise, and complex model reflection interference.  Studied were both problems induced by these complexities, and the advantages introduced by multiple extra data associated with the multiple reflectors (times and reflection strength), multiple stacks, differential tuning of the different stack bandwidths, and prior model assumptions.  The results confirm the significant value of multi-component Bayesian inversion (including PS data) and the feasibility of the detection of ductile fraction of the objective shales.


%


%

%

\end{document}